\documentclass[aps,twocolumn,showpacs,preprintnumbers,amsmath,amssymb,prl,nofootinbib,10pt]{revtex4-1}
\pdfoutput=1
\usepackage{pdfpages} 
\usepackage{graphicx}
\usepackage{dcolumn}
\usepackage[tight]{subfigure}
\usepackage{amsmath}
\usepackage{verbatim}
\usepackage{color}
\usepackage{bm} 
\usepackage{bbm}
\usepackage{natbib}
\usepackage{xspace}
\usepackage{marginnote}
\usepackage{mathtools}
\usepackage{dsfont}
\usepackage{hyperref} \hypersetup{colorlinks=true,linktoc=all,linkcolor=blue,breaklinks=true,citecolor=blue,urlcolor=blue}
\usepackage{etoolbox}
\robustify{\uparrow}
\robustify{\downarrow}
\robustify{\sum}
\robustify{\int}
\robustify{\nonumber}
\robustify{\cite}
\robustify{\footnote}
\newrobustcmd{\Figure}[2]{
  \begin{figure}[ht]
    \includegraphics[width=1.0\linewidth]{#1}
    \caption{#2}
  \end{figure}
}
\expandafter\newrobustcmd\csname Figure2\endcsname[3]{
\begin{figure}[ht]
  \includegraphics[width=0.9\linewidth]{#1}
  \\
  \includegraphics[width=0.9\linewidth]{#2}
  \caption{#3}
\end{figure}
}
\renewrobustcmd{\Re}{{\text{Re}}}
\renewrobustcmd{\Im}{{\text{Im}}}
\newrobustcmd{\eff}{\text{eff}} 
\newrobustcmd{\dagtot}{{\dag_\tot}}
\newrobustcmd{\dagres}{{\dag_\res}}
\newrobustcmd{\Ttot}{{\text{T}_\tot}}
\newrobustcmd{\Tres}{{\text{T}_\res}}
\newrobustcmd{\T}{{\text{T}}}
\newrobustcmd{\tot}{\text{tot}}
\newrobustcmd{\tun}{\text{T}}
\newrobustcmd{\res}{\text{R}}
\newrobustcmd{\un}{\text{i}}   
\newrobustcmd{\In}{\text{0}}   
\newrobustcmd{\state}{inverted stationary state\xspace}   
\newrobustcmd{\K}{\mathcal{K}}
\newrobustcmd{\D}{\mathcal{I}}
\renewrobustcmd{\P}{\mathcal{P}}   
\newrobustcmd{\W}{\tilde{W}}
\newrobustcmd{\Temp}{T}
\newrobustcmd{\dual}[1]{\bar{#1}}
\newrobustcmd{\one}{\mathds{1}}
\newrobustcmd{\ket}[1]{|#1\rangle}
\newrobustcmd{\bra}[1]{\langle#1|}
\newrobustcmd{\brkt}[1]{\langle #1 \rangle}
\newrobustcmd{\braket}[2]{\langle #1 | #2 \rangle}
\newrobustcmd{\Ket}[1]{\bm{|}#1\bm{)}}
\newrobustcmd{\Bra}[1]{\bm{(}#1\bm{|}}
\newrobustcmd{\Braket}[2]{\bm{(}#1\bm{|}#2\bm{)}}
\newrobustcmd{\Brkt}[1]{\bm{(} #1 \bm{)}}
\newrobustcmd{\op}[1]{\hat{#1}}
\DeclareMathOperator{\Tr}{Tr}
\newrobustcmd{\tr}{\underset{\res}{\Tr}}
\newrobustcmd{\tri}{\Tr_\res}
\newrobustcmd{\col}[4]{{\begin{bmatrix}#1 \\ #2 \\ #3 \\ #4 \end{bmatrix}}}
\newrobustcmd{\row}[4]{{\begin{bmatrix}#1 &  #2  & #3  & #4 \end{bmatrix}}}
\newrobustcmd{\suppmat}{\cite{Schulenborg15Suppmat}}
\newrobustcmd{\Eq}[1]{Eq.~(\ref{#1})}
\newrobustcmd{\eq}[1]{(\ref{#1})}
\newrobustcmd{\Fig}[1]{Fig.~\ref{#1}}
\newrobustcmd{\fig}[1]{\ref{#1}}
\newrobustcmd{\Figs}[1]{Figs.~\ref{#1}}
\newrobustcmd{\Sec}[1]{Sec.~\ref{#1}}
\newrobustcmd{\Ref}[1]{Ref.~[\onlinecite{#1}]}
\newrobustcmd{\Refs}[1]{Refs.~[\onlinecite{#1}]}
\usepackage[normalem]{ulem}

\begin{document}
\newrobustcmd{\suppmattitle}{}
\title{\suppmattitle
Fermion-parity duality
and energy relaxation
in interacting open systems
}
\author{J. Schulenborg$^{(1)}$}
\thanks{First and second author contributed equally.}
\author{R. B. Saptsov$^{(2,3)}$}
\thanks{First and second author contributed equally.}
\author{F. Haupt$^{(3,4)}$}
\author{J. Splettstoesser$^{(1)}$}
\author{M. R. Wegewijs$^{(2,3,5)}$}
\affiliation{
  (1) Department of Microtechnology and Nanoscience (MC2), Chalmers University of Technology, SE-41298 G{\"o}teborg, Sweden
  \\
  (2) Institute for Theory of Statistical Physics,
      RWTH Aachen, 52056 Aachen,  Germany
  \\
  (3) JARA- Fundamentals of Future Information Technology
  \\
  (4) JARA Institute for Quantum Information,
  RWTH Aachen, 52056 Aachen,  Germany
  \\
  (5) Peter Gr{\"u}nberg Institut,
      Forschungszentrum J{\"u}lich, 52425 J{\"u}lich,  Germany
}
\begin{abstract}
We study the transient heat current out of a confined electron system into a weakly coupled electrode in response to a voltage switch.
We show that the decay of the Coulomb interaction energy for this repulsive system exhibits signatures of electron-electron \emph{attraction}, and is governed by an interaction-independent rate.
This can only be understood from a general duality that relates the non-unitary evolution of a quantum system
to that of a dual model with inverted energies. Deriving from the fermion-parity superselection postulate, this duality applies to a large class of open systems.
\end{abstract}

\maketitle

Energy decay due to heat currents is of key importance in the continued downscaling of electronic devices~\cite{Schwab00}.
The quantum~\cite{Scully03,Bermudez13,Abah12,Whitney14a,Segal15} and interaction effects~\cite{Kubala08,Lotze12,Gergs15a} that arise on the nanoscale
 give rise to new possibilities~\cite{Giazotto06,Sanchez11,Jordan13,Sothmann14a}
 and motivate both fundamental~\cite{Koski14} and application oriented~\cite{Lotze12,Abah12,Brantut13,Bermudez13,Sothmann14a} studies on quantum heat-engines, possibly realized in, e.g., cold atoms, trapped ions, or quantum dots.  The successful control and exploitation of heat in nanodevices requires both a fundamental understanding and the practical ability to detect and manipulate \emph{few-electron} heat currents.
 Under stationary conditions, progress has been achieved  using various approaches~\cite{Scheibner05,Giazotto12,Svensson13}, including heat transfer through molecular-scale devices~\cite{LeeKim13} with electrostatic gating~\cite{Kim14}. However, any device is eventually adjusted by some external agent that provokes a time-dependent response. In the context of electronic heat currents, this raises a very basic question that, despite recent promising theoretical \cite{Moskalets09,Esposito10,Lim13,Juergens13,Battista13,Ludovico14,Zhou15} and experimental~\cite{Fletcher13,Ubbelohde15,Gasparinetti15} studies, has not been answered so far: how does a small electron system, typically governed by a strong level-quantization and Coulomb interaction, dissipate \emph{in time} its stored energy into a coupled electronic bath?
\begin{figure}[t]  
	\includegraphics[width=\linewidth]{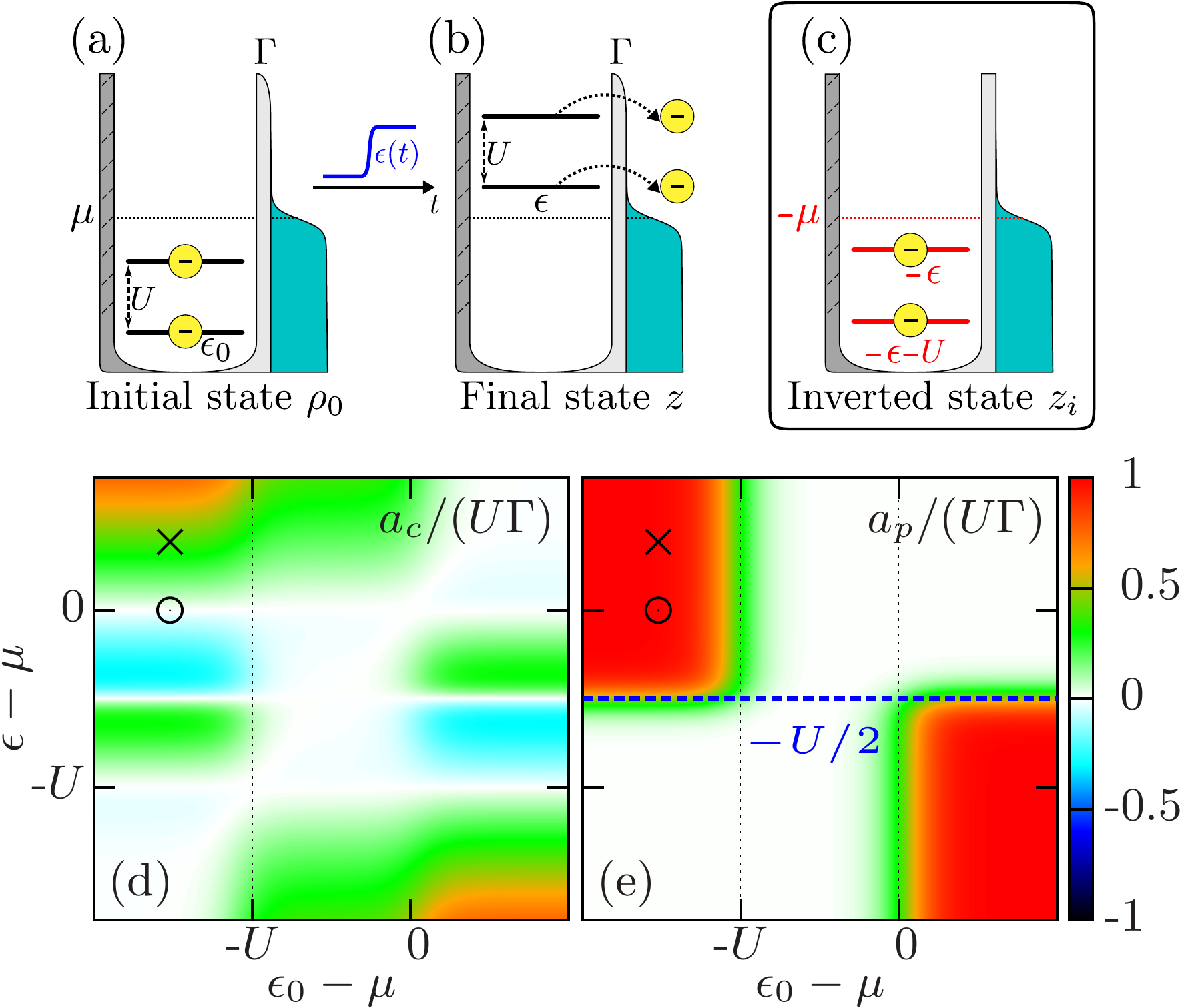}
	\caption{
                (a,b): Quantum dot with charging energy $U$ subject to a gate-voltage switch.
                The instant level shift from $\epsilon_\In$  to $\epsilon$ 
                causes two electrons to be sequentially expelled. 
		(c): Inverted stationary state for the \emph{dual} model obtained from the final state (b) by inverting all energies, as prescribed by \Eq{eq:duality}.
		(d,e): Amplitude of the 
		charge mode, $a_c$
		and of the 
		parity mode, $a_p$ 
		in the time-dependent heat current $I_Q(t)= a_c e^{-\gamma_c t} + a_p e^{-\gamma_p t}$, 
		plotted
                versus the initial ($\epsilon_\In$) and final ($\epsilon$) level position for $\Temp = 0.1 U \gg \Gamma$.
		The black crosses 
		mark the switch in (a,b), the circles a switch to $\epsilon = \mu$.
	}
	\label{fig:1}  
\end{figure}\par

The essence of time-dependent transport in such systems is already captured by the simple model sketched in \Figs{fig:1}(a,b). Here, an instant energy shift of a single electronic orbital in a quantum dot leads to a time-dependent charge current $I_N(t)$~\cite{Buettiker93,Buettiker94} and heat current $I_Q(t)$ into a tunnel-coupled electrode.
In the weak coupling regime, expressions for these currents can be calculated straightforwardly, and
in the case of the transient charge current $I_N(t)$ also
allow for an intuitive physical understanding~\cite{Splettstoesser10}.
This is, however, not the case for the \emph{heat current} $I_Q(t)= a_c e^{-\gamma_c t} + a_p e^{-\gamma_p t}$.
Compared to the charge current $I_N(t) \propto e^{-\gamma_c t}$, the heat current contains a second decay mode. The mere presence of this mode
can be expected: it originates from the dissipation of the Coulomb energy. However, what is quite remarkable is that its rate $\gamma_p$ turns out to be completely independent of the interaction strength $U$~\cite{Contreras12,Saptsov12a,Saptsov14a,Schulenborg14a} -- despite entering the heat current only as a consequence of the interaction.
Even more surprisingly, as indicated by the blue dashed line in \Fig{fig:1}(e), both excitation amplitudes, $a_c$ and $a_p$, show an abrupt change at an energy typically associated with electron-electron \emph{attraction}~\cite{Anderson75}, even though we are dealing with a system governed by \emph{repulsive} interaction.
As we will show, both surprising effects have a fundamental origin
and they can occur in a broad class of systems.\par

The difficulty in explaining the counter-intuitive observations in the heat current ultimately stems from the non-unitarity of the dynamics of open quantum systems. In closed systems, the hermiticity of the Hamiltonian implies that its left and right eigenvectors are simply adjoint to each other. Thanks to this property, the amplitude with which an energy eigenstate enters the full time evolution of the system is simply given by the overlap between this eigenstate and the initially prepared state.
For dissipative open systems, a similar and equally insightful relation for the amplitudes of the decay modes is not known. Yet, such a relation would be extremely valuable,
not only for the understanding of the transient heat-current in \Fig{fig:1}, but also for a broad class of other dynamical problems, ranging from qubit dynamics to molecular electron-transfer.\par

In this Rapid Communication, we identify a duality relation between decay modes and amplitudes. This duality applies to \emph{all open} fermionic systems that can be modeled by a Hamiltonian $H^\tot=H+H^\res+H^\tun$
with an arbitrary local Hamiltonian $H$ and
(a) effectively non-interacting, wide-band fermionic reservoirs $H^\res$,
 with a coupling $H^\tun$ that is
(b) bilinear in the fermion fields and
(c) energy-independent, but otherwise arbitrarily strong.
Under these very general
assumptions, the duality emerges
mainly as a consequence of the fermion-parity superselection postulate of quantum  mechanics~\cite{Wick52,Aharonov67,Streaterbook,Bogolubov89}, by which states with even and odd fermion number cannot be superposed. Applied to the system shown in \Fig{fig:1}, it explains in full detail the poorly understood heat decay.\par

\emph{Mode-amplitude duality for open systems.} An open quantum system is most naturally described in terms of its reduced density matrix $\rho(t)$,  whose equation of motion is $\partial_t\rho(t) = -i [H,\rho(t)] +\int_0^t dt' \mathcal{W}(t-t') \rho(t')$ [with $\hbar = k_\mathrm{B} = e = 1$]. The kernel $\mathcal{W}$ takes into account the coupling to the external reservoirs~\cite{Timm08,Cohen13a,Karlewski14,Schoeller09b}
that causes $\rho(t)$ to decay. Introducing the Laplace transform of $\mathcal{W}$, $W(\omega) =\int_0^\infty dt e^{i\omega t}\mathcal{W}(t)$, the decay dynamics can be expressed in terms of the frequency($\omega$)-dependent right eigenvectors of $W(\omega)$, the \emph{decay modes}. The decay amplitude for each mode is determined by the overlap between the initial state $\rho_0$ and the left eigenvector of $W(\omega)$ for the same eigenvalue, the \emph{amplitude covector}. Since $W$ is non-Hermitian, left and right eigenvectors are not simply each others' adjoint.  However, under the very general assumptions stated above, we can prove that the vectors are still linked by the duality relation
\begin{align}
    W(\omega; H, H^\tun, \{\mu\})^\dag
    = 
-\Gamma + \P W(\dual{\omega}; \dual{H},\dual{H}^{\tun},\{\dual{\mu}\})
\P. \label{eq:relation}
\end{align}
The physical consequences of \Eq{eq:relation} arise from the three  operations it involves:
(i)
a constant shift by $\Gamma$ -- the lumped sum of constant couplings characterizing $H^\tun$;
(ii)
the fundamental transformation $\P$, which multiplies an operator by the fermion-parity operator $(-\one)^N:=e^{i \pi N}$ with local fermion-number operator $N$;
(iii)
a parameter substitution in the original model, which constructs a dual model with inverted local energies $\dual{H} :=-H$, a dual coupling $\dual{H}^{\tun} := i H^\tun$
to reservoirs with dual chemical potentials $\dual{\mu} := -\mu$ at dual frequency $\dual{\omega}:=i\Gamma - \omega^{*}$, but with same Hamiltonian $H^\res$ and temperature $T$.
The duality \Eq{eq:relation} therefore links in a nontrivial way the left eigenvectors of $W$ to those of $W^\dag$, which are in turn the right eigenvectors of $W$ \emph{for the dual model}.
The physics behind the derivation of \Eq{eq:relation} is summarized at the end of the paper, and the proof is given in~\suppmat{}. In the following, we apply the duality to the transient dynamics of the system depicted in \Figs{fig:1}(a,b), and show that it provides the key insight to interpret the remarkable features of the heat current.\par

\emph{Transient dynamics.}
The system of interest is a spin-degenerate single-level quantum dot with Hamiltonian $H = \epsilon N + U N_\uparrow N_\downarrow$, where $\epsilon$ is the tuneable level position and $U$ the local interaction, see \Fig{fig:1}. Here, $N = N_\uparrow + N_\downarrow$ is the occupation operator, with $N_\sigma = d_\sigma^\dagger d_\sigma$ and $d_\sigma$ the field operators
of the dot electrons ($\sigma = \uparrow,\downarrow$). The dot is tunnel-coupled by $H^\tun = \sum_{k,\sigma}\tau_{k,\sigma}c_{k\sigma}^\dagger d_\sigma + \mathrm{H.c. }$  to a single noninteracting electrode $H^\res = \sum_{k,\sigma}\epsilon_{k,\sigma}c_{k \sigma}^{\dagger} c_{k\sigma}$. 
Here, $c_{k\sigma}$ is the field operator for reservoir electrons with  spin $\sigma$, orbital index $k$, and energy $\epsilon_{k,\sigma}$, while $\tau_{k,\sigma}$ is the tunnel amplitude. The grand-canonical reservoir state is
 $\rho^\res=e^{-(H^\res-\mu N^\res)/T}/\Tr_\res e^{-(H^\res-\mu N^\res)/T}$, where $N^\res = \sum_{k,\sigma} c_{k\sigma}^\dagger c_{k\sigma}$. In the wide-band limit, the tunnel coupling is characterized by $\Gamma_\sigma = 2\pi\sum_k\delta(\omega- \epsilon_{k})|\tau_{k,\sigma}|^2$, assumed $\omega$-independent, with  total strength $\Gamma := \sum_\sigma \Gamma_\sigma$ and $\Gamma_\uparrow = \Gamma_\downarrow = \Gamma/2$ in our spin-independent case.\par
 
We consider the regime of high-temperature and weak-coupling to the electrode  $\Gamma \ll T$, and the response to a sudden shift of the level position $\epsilon_{0}\to \epsilon$ at $t=0$, due to, e.g., a switch of the gate-voltage. 
Immediately after the level shift, the dot state equals the initial state $\rho_\In$ before the shift. At later times $t>0$,  the density operator $\rho(t)$ of the dot obeys the Born-Markov master equation
\begin{align}
  \partial_t\Ket{\rho(t)} \approx W \Ket{\rho(t)},\quad
  \quad W:=\lim_{\omega \to i0} W(\omega),
  \label{eq:master}
\end{align}
where $W= \sum_{i j} W_{ij}\Ket{i}\Bra{j}$ for $i,j=0,1,2$. Here and below, we write an operator $x$ as $\Ket{x} := x$, and
its covector acting on an argument $\bullet$ as $\Bra{x} \bullet := \Tr x^\dag \bullet $~\cite{Breuer}.
The basis vectors denote unit-trace physical-state operators~\footnote{The empty state is $\ket{0}$, whereas $\ket{\sigma} := d^\dagger_\sigma\ket{0}$ and $\ket{2}:=d^\dagger_\uparrow d^\dagger_\downarrow\ket{0}$.} with 0, 1 and 2 electrons:
$\Ket{0}:=\ket{0}\bra{0}$,
$\Ket{1}:=\tfrac{1}{2}\sum_\sigma \ket{\sigma}\bra{\sigma}$,
$\Ket{2}:=\ket{2}\bra{2}$. The coefficients $W_{ij}$, given in \suppmat{}, represent the golden-rule transition rates between the different electronic states (see, e.g., \cite{Contreras12}).\par

For $t > 0$, a standard way of solving \Eq{eq:master} proceeds by expanding $\rho(t)$ in the right eigenvectors of $W$ -- the decay modes -- and obtaining their coefficients from the corresponding left eigenvectors -- the amplitude covectors:
\begin{align}
	\Ket{\rho(t)} &= e^{W t} \Ket{\rho_0}\label{eq:rhot}\\
	&=  \Ket{z}\Braket{z'}{\rho_\In} + e^{-\gamma_{p}t} \Ket{p}  \Braket{p'}{\rho_\In} + e^{-\gamma_{c}t} \Ket{c}  \Braket{c'}{\rho_\In}.\notag
\end{align}
We denote the eigenvalues of $W$ by $-\gamma_x$, where $\gamma_x$ is a positive decay rate for $x=c,p$. 
Since the system reaches a unique stationary state $\lim_{t\to \infty}\Ket{\rho(t)}=\Ket{z}$,  one eigenvalue of $W$ is $0$.\par

To explicitly calculate the rates and vectors in \Eq{eq:rhot}, one can now straightforwardly determine the left and right eigenvectors of the 3 $\times$ 3 matrix representing $W$. However, this does not give
any systematic, \emph{physical} insight into how the excitation amplitudes and rates of the decay modes are related. As we illustrate in the following, the duality \Eq{eq:relation} does provide such insight. 
In the Born-Markov approximation, we expand \Eq{eq:relation} to linear order in $\Gamma$, and take the limit $\omega, i\Gamma \to i0$ in the frequency arguments.
This yields~\footnote{Since $W \propto \Gamma \propto (H^\tun)^2$, the dual coupling $H^\tun \to i{H}^\tun$ effectively inverts the sign of $W$.} $W^\dag = -\Gamma  -\P \D W \D \P$,
where $\D$ denotes the parameter substitution that constructs the dual model: $\mu \to -\mu$, $\epsilon \to -\epsilon$ and, importantly, $U \to -U$. For a known mode $\Ket{x}=x$ with decay rate $\gamma_x$, application of this relation to $\P \D \Ket{x}$ determines an amplitude covector $\Bra{y'}\bullet$ with
\begin{align}
  y' = (-\one)^N  \D  x \, \text{ and rate  }\, \gamma_y = \Gamma - \D \gamma_x.\label{eq:duality}
\end{align}
The key result~\Eq{eq:duality} \emph{cross-links} left and right eigenvectors with different rates. For our model, it allows to physically relate all the rates and vectors in \Eq{eq:rhot}.
First, the zero eigenvector of $W$ is simply the trace-normalized stationary equilibrium state
$\Ket{z} = e^{-(H - \mu N)/T}/\mathcal{Z}$ of the quantum dot with $\mathcal{Z}(\epsilon,U,\mu)=\Tr e^{-(H - \mu N)/T}$.
Its amplitude in \Eq{eq:rhot} is $\Braket{z'}{\rho_0} = 1$, since probability conservation ($\Tr \rho(t) =1$) for all times $t$ requires that the left eigenvector associated with the zero eigenvalue is 
$\Bra{z'} = \Bra{\one}$. The mere existence of this left zero eigenvector implies by duality \eq{eq:duality} that there is a maximal~\suppmat{}  rate $\gamma_p = \Gamma$  which depends only on the bare tunnel coupling, and whose mode is given by the fermion-parity operator:
\begin{align}
 \Ket{p} = \Ket{(-\one)^N} \text{ with } \gamma_p = \Gamma
  .
  \label{eq:ketp}
\end{align}
Analogously, the parity amplitude-covector is found by applying \Eq{eq:duality} to the stationary mode $\Ket{z}$, giving
\begin{align}
 \Bra{p'} = \Bra{(-\one)^N z_\un}  \text{ with } \gamma_p = \Gamma
  .
\end{align}
It contains $z_\un=\D z = z(-\epsilon,-U,-\mu)$,
the stationary state of the dual quantum-dot model with attractive interaction obtained by the parameter substitution $\D$, i.e, by inverting the energies as illustrated in \Fig{fig:1}(c).
Therefore, we call $\Ket{z_\un}$ the \emph{\state}. Finally, the charge-decay rate is self-dual, $\gamma_c   = \Gamma - \D\gamma_c = \tfrac{1}{2}\Gamma \{f^{+}(\epsilon) + f^{-}(\epsilon+U)\}$~\cite{Splettstoesser10} with $f^{\pm}(\epsilon)=(e^{\pm (\epsilon-\mu)/T}+1)^{-1}$. Its amplitude and mode (see~\suppmat{}) are thus also connected by the fermion-parity and the dual model: $\Bra{c'}=2[\P \D \Ket{c} ]^\dag$.\par

The existence of a decay mode with a rate $\Gamma$ that surprisingly only depends on the bare coupling was pointed out in previous works~\cite{Splettstoesser10,Contreras12,Saptsov12a,Saptsov14a,Schulenborg14a}. However, it was not understood where the independence from all physical attributes except the ``interface'' property $\Gamma$ stems from. The duality \eq{eq:duality} sheds an entirely new light onto this problem.
It shows that the constant decay rate $\Gamma$ is fixed by the fundamental requirement of probability conservation, via the duality based on the fermion-parity superselection principle.
Moreover, the duality relates the complete decay dynamics to the \emph{stationary state} $\Ket{z_{\un}}$ of the attractive dual model, as seen most explicitly in $\Bra{p'} = \Bra{(-\one)^Nz_{\un}}$. It is this relation to the dual model which dictates the amplitudes in the heat decay.\par

\emph{Heat decay.} We now apply the results derived above to study the transient heat current out of the dot after a sudden switch of the level position $\epsilon_\In \to \epsilon$. In the Born-Markov limit, we can evaluate this heat current~\suppmat{} as $I_Q(t)  = -\partial_t\Braket{H-\mu N}{\rho(t)}$, with $\rho(t)$ given in \Eq{eq:rhot}, and all its ingredients obtained by the duality \Eq{eq:duality}.
This way, one not only finds the announced double-exponential form of the heat current,
 $I_Q(t) = a_c e^{-\gamma_c t} + a_p e^{-\gamma_p t}$.
Importantly, the amplitudes
\begin{align}
  a_c  =& \left[ \epsilon-\mu + \tfrac{1}{2}(2-N_\un) U \right](N_\In - N_z)\gamma_c,\label{eq:ac}\\
  a_p  =& U\left[\tfrac{1}{2}(N_\un-1)(N_\In-1) + \tfrac{1}{4}(p_\un+p_\In)\right]\gamma_p\label{eq:ap},
\end{align}
can now also be expressed entirely in terms of quantities with a clear physical meaning: assuming that the dot is initially in the stationary state $\Ket{\rho_\In}=\Ket{z(\epsilon_\In)}$ with initial level position $\epsilon_{0}$,
the occupation number of the dot $N_\In = \Braket{N}{\rho_\In}$ and the parity $p_\In = \Braket{(-\one)^N}{\rho_\In}$ in the initial state carry all dependence on $\epsilon_\In$, as expected.
However, the dependence on the final level position $\epsilon$ enters not only through the stationary occupation $N_z = \Braket{N}{z}$, but also through the occupation $N_{\un}=\Braket{N}{z_{\un}}$ and parity $p_\un = \Braket{(-\one)^N}{z_{\un}}$ of the \emph{\state}, governed by attractive interactions. To illustrate the consequence of this dependence, we plot the amplitudes $a_{c}$, $a_{p}$ in \Figs{fig:1}(d,e) for a level switch  $\epsilon_{0} \to \epsilon$.
Most prominently, both $a_{c}$ and $a_{p}$ exhibit a very sharp, large change when tuning the \emph{final level} through $\epsilon-\mu=-U/2$. As revealed by Eqs.~\eq{eq:ac}-\eq{eq:ap}, the reason for this jump is that all terms in the heat current which relate to the Coulomb energy dissipation, i.e., the parity amplitude $a_p$ and the correction $\propto U$ in $a_c$, are governed by the dual stationary occupation number $N_{\un}$.
An attractive interaction, which here enters through the dual model, is well known~\cite{Anderson75} to force $N_{\un}$ to jump directly from $0$ to $2$ at $\epsilon-\mu = -U/2$, avoiding $N_\un=1$ and keeping an even parity $p_{\un} = +1$.
By contrast, a sweep of the \emph{initial level} $\epsilon_\In-\mu$ causes the initial charge to traverse $N_\In =0,1,2$, and the parity to alternate, $p_\In=1,-1,1$. This yields sharp changes of $a_p$ and $a_c$ at the expected energies for a repulsive system, the two Coulomb resonances $\epsilon_\In-\mu =0$ or $-U$. We stress that while the end result for $I_Q(t)$ also follows
from a straightforward calculation of left- and right- eigenvectors of $W$, it does not lead to an equally concise, physically-motivated form of $a_p$ and $a_c$. Most importantly, the peculiar $\epsilon$ dependence of these amplitudes is only revealed by the duality.\par

Another notable feature of \Figs{fig:1}(d,e) is that  the parity mode dominates $I_Q(t)$ whenever it is excited: in the red areas in \Fig{fig:1}(e), its amplitude assumes the constant, maximal value $a_p = \Gamma U $, whereas $|a_c| \lesssim \tfrac{1}{2} \Gamma U$~\suppmat{} in \Fig{fig:1}(d). This offers an interesting possibility of experimentally accessing the decay dynamics of the parity mode. 
Measuring $\gamma_p$ is a non-trivial task, since the parity mode $\Ket{(-\one)^N}$ does not enter single-particle observables like the average charge current, which indeed decays at a single rate: $I_N(t)  = -\partial_t\Braket{N}{\rho(t)} = (N_\In - N_z) \,  \gamma_c \, e^{-\gamma_c t}$. Methods to detect $\gamma_{p}$ have been proposed in Refs.~\cite{Contreras12,Schulenborg14a}, by coupling the dot to a quantum point-contact or a sensor quantum-dot. \Figs{fig:1}(d,e) show that the heat current provides a very natural and more direct way to gain this information.
In fact, $I_Q(t)$ can be obtained using a pump-probe scheme by extending well-established techniques of mesoscopic charge detection~\cite{Fujisawa03rev,Roessler13}.
By measuring the rise of the stationary electrode temperature as a function of the delay time, one can extract the full \emph{time-resolved} heat current. Simultaneous measurement of the charge current then allows
to demonstrate -- without fitting -- the predicted constancy of the rate $\gamma_p = \Gamma$ as well as the attractive-interaction signature of $a_p$, as sketched in \suppmat.\par

An important insight for this kind of experiments is provided again by the duality and is illustrated in \Figs{fig:1}(a-c).
For $U \gg T$, the amplitude $a_p \approx \Braket{z_\un}{\rho_\In} U\Gamma$ essentially equals the overlap of the initial state with the inverted stationary state.
To infer which switch excites the parity mode, one can thus proceed as follows: given the targeted final level position  $\epsilon$, one calculates the stationary state of the attractive dual system, $\Ket{z_\un(\epsilon)}$, and chooses the initial level position $\epsilon_\In$ such that the initially prepared state $\Ket{\rho_\In} = \Ket{z(\epsilon_\In)}$ has the same occupation as $\Ket{z_\un(\epsilon)}$.\par 

In such measurements of the heat current, one can \emph{resolve in time} that two sequentially tunneling electrons carry equal charge but different energy.
(Only for $U = 0$, the heat current is simply ``tightly coupled'' to the particle current~\cite{Sothmann14a}, $I_Q \propto I_N$ since $a_p=0$.)
For example, for an initially doubly occupied dot and a switch expelling both electrons -- the situation at the cross in \Figs{fig:1}(d,e) -- the heat current simplifies to $I_Q(t) \approx (\epsilon-\mu) I_N(t) + U \Gamma e^{-\gamma_{p} t}$. While the excess orbital energy $\epsilon-\mu$  is carried by each of the two electrons tunneling out, the charging energy $U$ is dissipated already with the \emph{first} electron. Notably, when switching  $\epsilon$ to one of the two Coulomb resonances -- e.g., at the open circle in \Figs{fig:1}(d,e) -- the heat-current is even \emph{entirely} due to the tunneling of the first particle, taking place on the shortest, temperature independent time-scale $\Gamma^{-1}$.\par

\emph{General duality.}
We conclude by discussing the main physical principles behind the \emph{general} duality relation \Eq{eq:relation}. The derivation of \Eq{eq:relation} is technical but straightforward \footnote{\Eq{eq:relation} follows by combining Eqs. (E-7), (E-8), and (G-6) of \cite{Saptsov12a} with Eqs. (38), (65), and (101) of \cite{Saptsov14a}. The derivation is written out in~\suppmat{}, cf. Eq. (S-61) and (S-71).} if one uses the insights established in~\cite{Saptsov12a,Saptsov14a}. 
The main point is that in the wide-band limit, the best reference solution for a perturbative calculation of the dynamics is \emph{not} the uncoupled solution ($H^\tun=0$), but rather the exact solution of the coupled system in the limit of infinite temperature $T \to \infty$. With respect to this solution, the propagator $\Pi(t)$ for the density matrix of the system, $\rho(t)=\Pi(t)\rho_\In$, needs to be expanded in terms of only \emph{part} of the coupling, as a consequence of the fermion-parity superselection principle~\cite{Saptsov12a,Saptsov14a}.
What is crucial for the result \Eq{eq:relation} is that the adjoints of the operators occurring in this simpler expansion can always be expressed in parity operations~\suppmat{}. This allows to derive \emph{order-by-order} the time-propagator duality $\Pi(t; H,H^\tun,\{\mu\})^\dag = e^{-\Gamma t} \P \Pi(t; \dual{H}, \dual{H}^{\tun},\{\dual{\mu}\}) \P$. This is equivalent to \Eq{eq:relation} and holds under the general conditions stated in the introduction, i.e. it applies also to nontrivial low-temperature, strong-coupling, non-equilibrium regimes of complex multi-level fermionic systems. In the commonly used expansion about the uncoupled system ($H^\tun=0$), the duality remains completely elusive.
Extending the duality beyond the wide-band limit is challenging but seems possible using~\cite{Saptsov12a,Saptsov14a}.\par

We acknowledge discussions with
J. C. Cuevas,
N. Dittmann,
D. DiVincenzo,
M. Hell,
M. Pletyukhov,
T. Pl\"ucker,
R. Sanchez,
and financial support of 
DFG project SCHO 641/7-1 (R.B.S.),
the Swedish VR
and the Knut and Alice Wallenberg foundation (J.Sc., J. Sp.).
% \bibliography{cite,suppmatRef}

\begin{thebibliography}{50}%
\makeatletter
\providecommand \@ifxundefined [1]{%
 \@ifx{#1\undefined}
}%
\providecommand \@ifnum [1]{%
 \ifnum #1\expandafter \@firstoftwo
 \else \expandafter \@secondoftwo
 \fi
}%
\providecommand \@ifx [1]{%
 \ifx #1\expandafter \@firstoftwo
 \else \expandafter \@secondoftwo
 \fi
}%
\providecommand \natexlab [1]{#1}%
\providecommand \enquote  [1]{``#1''}%
\providecommand \bibnamefont  [1]{#1}%
\providecommand \bibfnamefont [1]{#1}%
\providecommand \citenamefont [1]{#1}%
\providecommand \href@noop [0]{\@secondoftwo}%
\providecommand \href [0]{\begingroup \@sanitize@url \@href}%
\providecommand \@href[1]{\@@startlink{#1}\@@href}%
\providecommand \@@href[1]{\endgroup#1\@@endlink}%
\providecommand \@sanitize@url [0]{\catcode `\\12\catcode `\$12\catcode
  `\&12\catcode `\#12\catcode `\^12\catcode `\_12\catcode `\%12\relax}%
\providecommand \@@startlink[1]{}%
\providecommand \@@endlink[0]{}%
\providecommand \url  [0]{\begingroup\@sanitize@url \@url }%
\providecommand \@url [1]{\endgroup\@href {#1}{\urlprefix }}%
\providecommand \urlprefix  [0]{URL }%
\providecommand \Eprint [0]{\href }%
\providecommand \doibase [0]{http://dx.doi.org/}%
\providecommand \selectlanguage [0]{\@gobble}%
\providecommand \bibinfo  [0]{\@secondoftwo}%
\providecommand \bibfield  [0]{\@secondoftwo}%
\providecommand \translation [1]{[#1]}%
\providecommand \BibitemOpen [0]{}%
\providecommand \bibitemStop [0]{}%
\providecommand \bibitemNoStop [0]{.\EOS\space}%
\providecommand \EOS [0]{\spacefactor3000\relax}%
\providecommand \BibitemShut  [1]{\csname bibitem#1\endcsname}%
\let\auto@bib@innerbib\@empty
%</preamble>
\bibitem [{\citenamefont {Schwab}\ \emph {et~al.}(2000)\citenamefont {Schwab},
  \citenamefont {Henriksen}, \citenamefont {Worlock},\ and\ \citenamefont
  {Roukes}}]{Schwab00}%
  \BibitemOpen
  \bibfield  {author} {\bibinfo {author} {\bibfnamefont {K.}~\bibnamefont
  {Schwab}}, \bibinfo {author} {\bibfnamefont {E.}~\bibnamefont {Henriksen}},
  \bibinfo {author} {\bibfnamefont {J.}~\bibnamefont {Worlock}}, \ and\
  \bibinfo {author} {\bibfnamefont {M.}~\bibnamefont {Roukes}},\ }\href
  {\doibase 10.1038/35010065} {\bibfield  {journal} {\bibinfo  {journal}
  {Nature}\ }\textbf {\bibinfo {volume} {404}},\ \bibinfo {pages} {974}
  (\bibinfo {year} {2000})}\BibitemShut {NoStop}%
\bibitem [{\citenamefont {Scully}\ \emph {et~al.}(2003)\citenamefont {Scully},
  \citenamefont {Zubairy}, \citenamefont {Agarwal},\ and\ \citenamefont
  {Walther}}]{Scully03}%
  \BibitemOpen
  \bibfield  {author} {\bibinfo {author} {\bibfnamefont {M.~O.}\ \bibnamefont
  {Scully}}, \bibinfo {author} {\bibfnamefont {M.~S.}\ \bibnamefont {Zubairy}},
  \bibinfo {author} {\bibfnamefont {G.~S.}\ \bibnamefont {Agarwal}}, \ and\
  \bibinfo {author} {\bibfnamefont {H.}~\bibnamefont {Walther}},\ }\href
  {\doibase 10.1126/science.1078955} {\bibfield  {journal} {\bibinfo  {journal}
  {Science}\ }\textbf {\bibinfo {volume} {299}},\ \bibinfo {pages} {862}
  (\bibinfo {year} {2003})}\BibitemShut {NoStop}%
\bibitem [{\citenamefont {Bermudez}\ \emph {et~al.}(2013)\citenamefont
  {Bermudez}, \citenamefont {Bruderer},\ and\ \citenamefont
  {Plenio}}]{Bermudez13}%
  \BibitemOpen
  \bibfield  {author} {\bibinfo {author} {\bibfnamefont {A.}~\bibnamefont
  {Bermudez}}, \bibinfo {author} {\bibfnamefont {M.}~\bibnamefont {Bruderer}},
  \ and\ \bibinfo {author} {\bibfnamefont {M.~B.}\ \bibnamefont {Plenio}},\
  }\href {\doibase 10.1103/PhysRevLett.111.040601} {\bibfield  {journal}
  {\bibinfo  {journal} {Phys. Rev. Lett.}\ }\textbf {\bibinfo {volume} {111}},\
  \bibinfo {pages} {040601} (\bibinfo {year} {2013})}\BibitemShut {NoStop}%
\bibitem [{\citenamefont {Abah}\ \emph {et~al.}(2012)\citenamefont {Abah},
  \citenamefont {Ro\ss{}nagel}, \citenamefont {Jacob}, \citenamefont {Deffner},
  \citenamefont {Schmidt-Kaler}, \citenamefont {Singer},\ and\ \citenamefont
  {Lutz}}]{Abah12}%
  \BibitemOpen
  \bibfield  {author} {\bibinfo {author} {\bibfnamefont {O.}~\bibnamefont
  {Abah}}, \bibinfo {author} {\bibfnamefont {J.}~\bibnamefont {Ro\ss{}nagel}},
  \bibinfo {author} {\bibfnamefont {G.}~\bibnamefont {Jacob}}, \bibinfo
  {author} {\bibfnamefont {S.}~\bibnamefont {Deffner}}, \bibinfo {author}
  {\bibfnamefont {F.}~\bibnamefont {Schmidt-Kaler}}, \bibinfo {author}
  {\bibfnamefont {K.}~\bibnamefont {Singer}}, \ and\ \bibinfo {author}
  {\bibfnamefont {E.}~\bibnamefont {Lutz}},\ }\href {\doibase
  10.1103/PhysRevLett.109.203006} {\bibfield  {journal} {\bibinfo  {journal}
  {Phys. Rev. Lett.}\ }\textbf {\bibinfo {volume} {109}},\ \bibinfo {pages}
  {203006} (\bibinfo {year} {2012})}\BibitemShut {NoStop}%
\bibitem [{\citenamefont {Whitney}(2014)}]{Whitney14a}%
  \BibitemOpen
  \bibfield  {author} {\bibinfo {author} {\bibfnamefont {R.~S.}\ \bibnamefont
  {Whitney}},\ }\href {\doibase 10.1103/PhysRevLett.112.130601} {\bibfield
  {journal} {\bibinfo  {journal} {Phys. Rev. Lett.}\ }\textbf {\bibinfo
  {volume} {112}},\ \bibinfo {pages} {130601} (\bibinfo {year}
  {2014})}\BibitemShut {NoStop}%
\bibitem [{\citenamefont {Taylor}\ and\ \citenamefont {Segal}(2015)}]{Segal15}%
  \BibitemOpen
  \bibfield  {author} {\bibinfo {author} {\bibfnamefont {E.}~\bibnamefont
  {Taylor}}\ and\ \bibinfo {author} {\bibfnamefont {D.}~\bibnamefont {Segal}},\
  }\href {\doibase 10.1103/PhysRevLett.114.220401} {\bibfield  {journal}
  {\bibinfo  {journal} {Phys. Rev. Lett.}\ }\textbf {\bibinfo {volume} {114}},\
  \bibinfo {pages} {220401} (\bibinfo {year} {2015})}\BibitemShut {NoStop}%
\bibitem [{\citenamefont {Kubala}\ \emph {et~al.}(2008)\citenamefont {Kubala},
  \citenamefont {K\"onig},\ and\ \citenamefont {Pekola}}]{Kubala08}%
  \BibitemOpen
  \bibfield  {author} {\bibinfo {author} {\bibfnamefont {B.}~\bibnamefont
  {Kubala}}, \bibinfo {author} {\bibfnamefont {J.}~\bibnamefont {K\"onig}}, \
  and\ \bibinfo {author} {\bibfnamefont {J.}~\bibnamefont {Pekola}},\ }\href
  {\doibase 10.1103/PhysRevLett.100.066801} {\bibfield  {journal} {\bibinfo
  {journal} {Phys. Rev. Lett.}\ }\textbf {\bibinfo {volume} {100}},\ \bibinfo
  {pages} {066801} (\bibinfo {year} {2008})}\BibitemShut {NoStop}%
\bibitem [{\citenamefont {Lotze}\ \emph {et~al.}(2012)\citenamefont {Lotze},
  \citenamefont {Corso}, \citenamefont {Franke}, \citenamefont {von Oppen},\
  and\ \citenamefont {Pascual}}]{Lotze12}%
  \BibitemOpen
  \bibfield  {author} {\bibinfo {author} {\bibfnamefont {C.}~\bibnamefont
  {Lotze}}, \bibinfo {author} {\bibfnamefont {M.}~\bibnamefont {Corso}},
  \bibinfo {author} {\bibfnamefont {K.~J.}\ \bibnamefont {Franke}}, \bibinfo
  {author} {\bibfnamefont {F.}~\bibnamefont {von Oppen}}, \ and\ \bibinfo
  {author} {\bibfnamefont {J.~I.}\ \bibnamefont {Pascual}},\ }\href {\doibase
  10.1126/science.1227621} {\bibfield  {journal} {\bibinfo  {journal}
  {Science}\ }\textbf {\bibinfo {volume} {338}},\ \bibinfo {pages} {779}
  (\bibinfo {year} {2012})}\BibitemShut {NoStop}%
\bibitem [{\citenamefont {Gergs}\ \emph {et~al.}(2015)\citenamefont {Gergs},
  \citenamefont {H\"orig}, \citenamefont {Wegewijs},\ and\ \citenamefont
  {Schuricht}}]{Gergs15a}%
  \BibitemOpen
  \bibfield  {author} {\bibinfo {author} {\bibfnamefont {N.~M.}\ \bibnamefont
  {Gergs}}, \bibinfo {author} {\bibfnamefont {C.~B.~M.}\ \bibnamefont
  {H\"orig}}, \bibinfo {author} {\bibfnamefont {M.~R.}\ \bibnamefont
  {Wegewijs}}, \ and\ \bibinfo {author} {\bibfnamefont {D.}~\bibnamefont
  {Schuricht}},\ }\href {\doibase 10.1103/PhysRevB.91.201107} {\bibfield
  {journal} {\bibinfo  {journal} {Phys. Rev. B}\ }\textbf {\bibinfo {volume}
  {91}},\ \bibinfo {pages} {201107} (\bibinfo {year} {2015})}\BibitemShut
  {NoStop}%
\bibitem [{\citenamefont {Giazotto}\ \emph {et~al.}(2006)\citenamefont
  {Giazotto}, \citenamefont {Heikkil\"a}, \citenamefont {Luukanen},
  \citenamefont {Savin},\ and\ \citenamefont {Pekola}}]{Giazotto06}%
  \BibitemOpen
  \bibfield  {author} {\bibinfo {author} {\bibfnamefont {F.}~\bibnamefont
  {Giazotto}}, \bibinfo {author} {\bibfnamefont {T.~T.}\ \bibnamefont
  {Heikkil\"a}}, \bibinfo {author} {\bibfnamefont {A.}~\bibnamefont
  {Luukanen}}, \bibinfo {author} {\bibfnamefont {A.~M.}\ \bibnamefont {Savin}},
  \ and\ \bibinfo {author} {\bibfnamefont {J.~P.}\ \bibnamefont {Pekola}},\
  }\href {\doibase 10.1103/RevModPhys.78.217} {\bibfield  {journal} {\bibinfo
  {journal} {Rev. Mod. Phys.}\ }\textbf {\bibinfo {volume} {78}},\ \bibinfo
  {pages} {217} (\bibinfo {year} {2006})}\BibitemShut {NoStop}%
\bibitem [{\citenamefont {S\'anchez}\ and\ \citenamefont
  {B\"uttiker}(2011)}]{Sanchez11}%
  \BibitemOpen
  \bibfield  {author} {\bibinfo {author} {\bibfnamefont {R.}~\bibnamefont
  {S\'anchez}}\ and\ \bibinfo {author} {\bibfnamefont {M.}~\bibnamefont
  {B\"uttiker}},\ }\href {\doibase 10.1103/PhysRevB.83.085428} {\bibfield
  {journal} {\bibinfo  {journal} {Phys. Rev. B}\ }\textbf {\bibinfo {volume}
  {83}},\ \bibinfo {pages} {085428} (\bibinfo {year} {2011})}\BibitemShut
  {NoStop}%
\bibitem [{\citenamefont {Jordan}\ \emph {et~al.}(2013)\citenamefont {Jordan},
  \citenamefont {Sothmann}, \citenamefont {S\'anchez},\ and\ \citenamefont
  {B\"uttiker}}]{Jordan13}%
  \BibitemOpen
  \bibfield  {author} {\bibinfo {author} {\bibfnamefont {A.~N.}\ \bibnamefont
  {Jordan}}, \bibinfo {author} {\bibfnamefont {B.}~\bibnamefont {Sothmann}},
  \bibinfo {author} {\bibfnamefont {R.}~\bibnamefont {S\'anchez}}, \ and\
  \bibinfo {author} {\bibfnamefont {M.}~\bibnamefont {B\"uttiker}},\ }\href
  {\doibase 10.1103/PhysRevB.87.075312} {\bibfield  {journal} {\bibinfo
  {journal} {Phys. Rev. B}\ }\textbf {\bibinfo {volume} {87}},\ \bibinfo
  {pages} {075312} (\bibinfo {year} {2013})}\BibitemShut {NoStop}%
\bibitem [{\citenamefont {Sothmann}\ \emph {et~al.}(2015)\citenamefont
  {Sothmann}, \citenamefont {S\'anchez},\ and\ \citenamefont
  {Jordan}}]{Sothmann14a}%
  \BibitemOpen
  \bibfield  {author} {\bibinfo {author} {\bibfnamefont {B.}~\bibnamefont
  {Sothmann}}, \bibinfo {author} {\bibfnamefont {R.}~\bibnamefont {S\'anchez}},
  \ and\ \bibinfo {author} {\bibfnamefont {A.~N.}\ \bibnamefont {Jordan}},\
  }\href {http://stacks.iop.org/0957-4484/26/i=3/a=032001} {\bibfield
  {journal} {\bibinfo  {journal} {Nanotechnology}\ }\textbf {\bibinfo {volume}
  {26}},\ \bibinfo {pages} {032001} (\bibinfo {year} {2015})}\BibitemShut
  {NoStop}%
\bibitem [{\citenamefont {Koski}\ \emph {et~al.}(2014)\citenamefont {Koski},
  \citenamefont {Maisi}, \citenamefont {Pekola},\ and\ \citenamefont
  {Averin}}]{Koski14}%
  \BibitemOpen
  \bibfield  {author} {\bibinfo {author} {\bibfnamefont {J.~V.}\ \bibnamefont
  {Koski}}, \bibinfo {author} {\bibfnamefont {V.~F.}\ \bibnamefont {Maisi}},
  \bibinfo {author} {\bibfnamefont {J.~P.}\ \bibnamefont {Pekola}}, \ and\
  \bibinfo {author} {\bibfnamefont {D.~V.}\ \bibnamefont {Averin}},\ }\href
  {\doibase 10.1073/pnas.1406966111} {\bibfield  {journal} {\bibinfo  {journal}
  {P. Natl. Acad. Sci. USA}\ }\textbf {\bibinfo {volume} {111}},\ \bibinfo
  {pages} {13786} (\bibinfo {year} {2014})}\BibitemShut {NoStop}%
\bibitem [{\citenamefont {Brantut}\ \emph {et~al.}(2013)\citenamefont
  {Brantut}, \citenamefont {Grenier}, \citenamefont {Meineke}, \citenamefont
  {Stadler}, \citenamefont {Krinner}, \citenamefont {Kollath}, \citenamefont
  {Esslinger},\ and\ \citenamefont {Georges}}]{Brantut13}%
  \BibitemOpen
  \bibfield  {author} {\bibinfo {author} {\bibfnamefont {J.-P.}\ \bibnamefont
  {Brantut}}, \bibinfo {author} {\bibfnamefont {C.}~\bibnamefont {Grenier}},
  \bibinfo {author} {\bibfnamefont {J.}~\bibnamefont {Meineke}}, \bibinfo
  {author} {\bibfnamefont {D.}~\bibnamefont {Stadler}}, \bibinfo {author}
  {\bibfnamefont {S.}~\bibnamefont {Krinner}}, \bibinfo {author} {\bibfnamefont
  {C.}~\bibnamefont {Kollath}}, \bibinfo {author} {\bibfnamefont
  {T.}~\bibnamefont {Esslinger}}, \ and\ \bibinfo {author} {\bibfnamefont
  {A.}~\bibnamefont {Georges}},\ }\href {\doibase 10.1126/science.1242308}
  {\bibfield  {journal} {\bibinfo  {journal} {Science}\ }\textbf {\bibinfo
  {volume} {342}},\ \bibinfo {pages} {713} (\bibinfo {year}
  {2013})}\BibitemShut {NoStop}%
\bibitem [{\citenamefont {Scheibner}\ \emph {et~al.}(2005)\citenamefont
  {Scheibner}, \citenamefont {Buhmann}, \citenamefont {Reuter}, \citenamefont
  {Kiselev},\ and\ \citenamefont {Molenkamp}}]{Scheibner05}%
  \BibitemOpen
  \bibfield  {author} {\bibinfo {author} {\bibfnamefont {R.}~\bibnamefont
  {Scheibner}}, \bibinfo {author} {\bibfnamefont {H.}~\bibnamefont {Buhmann}},
  \bibinfo {author} {\bibfnamefont {D.}~\bibnamefont {Reuter}}, \bibinfo
  {author} {\bibfnamefont {M.~N.}\ \bibnamefont {Kiselev}}, \ and\ \bibinfo
  {author} {\bibfnamefont {L.~W.}\ \bibnamefont {Molenkamp}},\ }\href {\doibase
  10.1103/PhysRevLett.95.176602} {\bibfield  {journal} {\bibinfo  {journal}
  {Phys. Rev. Lett.}\ }\textbf {\bibinfo {volume} {95}},\ \bibinfo {pages}
  {176602} (\bibinfo {year} {2005})}\BibitemShut {NoStop}%
\bibitem [{\citenamefont {Giazotto}\ and\ \citenamefont
  {Mart\'{i}nez-P\'{e}rez}(2012)}]{Giazotto12}%
  \BibitemOpen
  \bibfield  {author} {\bibinfo {author} {\bibfnamefont {F.}~\bibnamefont
  {Giazotto}}\ and\ \bibinfo {author} {\bibfnamefont {M.~J.}\ \bibnamefont
  {Mart\'{i}nez-P\'{e}rez}},\ }\href {\doibase 10.1038/nature11702} {\bibfield
  {journal} {\bibinfo  {journal} {Nature}\ }\textbf {\bibinfo {volume} {492}},\
  \bibinfo {pages} {401} (\bibinfo {year} {2012})}\BibitemShut {NoStop}%
\bibitem [{\citenamefont {Svensson}\ \emph {et~al.}(2013)\citenamefont
  {Svensson}, \citenamefont {Hoffmann}, \citenamefont {Nakpathomkun},
  \citenamefont {Wu}, \citenamefont {Xu}, \citenamefont {Nilsson},
  \citenamefont {S{\'a}nchez}, \citenamefont {Kashcheyevs},\ and\ \citenamefont
  {Linke}}]{Svensson13}%
  \BibitemOpen
  \bibfield  {author} {\bibinfo {author} {\bibfnamefont {S.~F.}\ \bibnamefont
  {Svensson}}, \bibinfo {author} {\bibfnamefont {E.~A.}\ \bibnamefont
  {Hoffmann}}, \bibinfo {author} {\bibfnamefont {N.}~\bibnamefont
  {Nakpathomkun}}, \bibinfo {author} {\bibfnamefont {P.~M.}\ \bibnamefont
  {Wu}}, \bibinfo {author} {\bibfnamefont {H.}~\bibnamefont {Xu}}, \bibinfo
  {author} {\bibfnamefont {H.~A.}\ \bibnamefont {Nilsson}}, \bibinfo {author}
  {\bibfnamefont {D.}~\bibnamefont {S{\'a}nchez}}, \bibinfo {author}
  {\bibfnamefont {V.}~\bibnamefont {Kashcheyevs}}, \ and\ \bibinfo {author}
  {\bibfnamefont {H.}~\bibnamefont {Linke}},\ }\href
  {http://stacks.iop.org/1367-2630/15/i=10/a=105011} {\bibfield  {journal}
  {\bibinfo  {journal} {New. J. Phys.}\ }\textbf {\bibinfo {volume} {15}},\
  \bibinfo {pages} {105011} (\bibinfo {year} {2013})}\BibitemShut {NoStop}%
\bibitem [{\citenamefont {Lee}\ \emph {et~al.}(2013)\citenamefont {Lee},
  \citenamefont {Kim}, \citenamefont {Jeong}, \citenamefont {Zotti},
  \citenamefont {Pauly}, \citenamefont {Cuevas},\ and\ \citenamefont
  {Reddy}}]{LeeKim13}%
  \BibitemOpen
  \bibfield  {author} {\bibinfo {author} {\bibfnamefont {W.}~\bibnamefont
  {Lee}}, \bibinfo {author} {\bibfnamefont {K.}~\bibnamefont {Kim}}, \bibinfo
  {author} {\bibfnamefont {W.}~\bibnamefont {Jeong}}, \bibinfo {author}
  {\bibfnamefont {L.~A.}\ \bibnamefont {Zotti}}, \bibinfo {author}
  {\bibfnamefont {F.}~\bibnamefont {Pauly}}, \bibinfo {author} {\bibfnamefont
  {J.}~\bibnamefont {Cuevas}}, \ and\ \bibinfo {author} {\bibfnamefont
  {P.}~\bibnamefont {Reddy}},\ }\href {http://dx.doi.org/10.1038/nature12183}
  {\bibfield  {journal} {\bibinfo  {journal} {Nature}\ }\textbf {\bibinfo
  {volume} {498}},\ \bibinfo {pages} {209} (\bibinfo {year}
  {2013})}\BibitemShut {NoStop}%
\bibitem [{\citenamefont {Kim}\ \emph {et~al.}(2014)\citenamefont {Kim},
  \citenamefont {Jeong}, \citenamefont {Kim}, \citenamefont {Lee},\ and\
  \citenamefont {Reddy}}]{Kim14}%
  \BibitemOpen
  \bibfield  {author} {\bibinfo {author} {\bibfnamefont {Y.}~\bibnamefont
  {Kim}}, \bibinfo {author} {\bibfnamefont {W.}~\bibnamefont {Jeong}}, \bibinfo
  {author} {\bibfnamefont {K.}~\bibnamefont {Kim}}, \bibinfo {author}
  {\bibfnamefont {W.}~\bibnamefont {Lee}}, \ and\ \bibinfo {author}
  {\bibfnamefont {P.}~\bibnamefont {Reddy}},\ }\href
  {http://dx.doi.org/10.1038/nnano.2014.209} {\bibfield  {journal} {\bibinfo
  {journal} {Nat. Nanotechnol.}\ }\textbf {\bibinfo {volume} {9}},\ \bibinfo
  {pages} {881} (\bibinfo {year} {2014})}\BibitemShut {NoStop}%
\bibitem [{\citenamefont {Moskalets}\ and\ \citenamefont
  {B\"uttiker}(2009)}]{Moskalets09}%
  \BibitemOpen
  \bibfield  {author} {\bibinfo {author} {\bibfnamefont {M.}~\bibnamefont
  {Moskalets}}\ and\ \bibinfo {author} {\bibfnamefont {M.}~\bibnamefont
  {B\"uttiker}},\ }\href {\doibase 10.1103/PhysRevB.80.081302} {\bibfield
  {journal} {\bibinfo  {journal} {Phys. Rev. B}\ }\textbf {\bibinfo {volume}
  {80}},\ \bibinfo {pages} {081302} (\bibinfo {year} {2009})}\BibitemShut
  {NoStop}%
\bibitem [{\citenamefont {Esposito}\ \emph {et~al.}(2010)\citenamefont
  {Esposito}, \citenamefont {Kawai}, \citenamefont {Lindenberg},\ and\
  \citenamefont {van~den Broeck}}]{Esposito10}%
  \BibitemOpen
  \bibfield  {author} {\bibinfo {author} {\bibfnamefont {M.}~\bibnamefont
  {Esposito}}, \bibinfo {author} {\bibfnamefont {R.}~\bibnamefont {Kawai}},
  \bibinfo {author} {\bibfnamefont {K.}~\bibnamefont {Lindenberg}}, \ and\
  \bibinfo {author} {\bibfnamefont {C.}~\bibnamefont {van~den Broeck}},\ }\href
  {http://stacks.iop.org/0295-5075/89/i=2/a=20003} {\bibfield  {journal}
  {\bibinfo  {journal} {EPL}\ }\textbf {\bibinfo {volume} {89}},\ \bibinfo
  {pages} {20003} (\bibinfo {year} {2010})}\BibitemShut {NoStop}%
\bibitem [{\citenamefont {Lim}\ \emph {et~al.}(2013)\citenamefont {Lim},
  \citenamefont {L\'opez},\ and\ \citenamefont {S\'anchez}}]{Lim13}%
  \BibitemOpen
  \bibfield  {author} {\bibinfo {author} {\bibfnamefont {J.~S.}\ \bibnamefont
  {Lim}}, \bibinfo {author} {\bibfnamefont {R.}~\bibnamefont {L\'opez}}, \ and\
  \bibinfo {author} {\bibfnamefont {D.}~\bibnamefont {S\'anchez}},\ }\href
  {\doibase 10.1103/PhysRevB.88.201304} {\bibfield  {journal} {\bibinfo
  {journal} {Phys. Rev. B}\ }\textbf {\bibinfo {volume} {88}},\ \bibinfo
  {pages} {201304} (\bibinfo {year} {2013})}\BibitemShut {NoStop}%
\bibitem [{\citenamefont {Juergens}\ \emph {et~al.}(2013)\citenamefont
  {Juergens}, \citenamefont {Haupt}, \citenamefont {Moskalets},\ and\
  \citenamefont {Splettstoesser}}]{Juergens13}%
  \BibitemOpen
  \bibfield  {author} {\bibinfo {author} {\bibfnamefont {S.}~\bibnamefont
  {Juergens}}, \bibinfo {author} {\bibfnamefont {F.}~\bibnamefont {Haupt}},
  \bibinfo {author} {\bibfnamefont {M.}~\bibnamefont {Moskalets}}, \ and\
  \bibinfo {author} {\bibfnamefont {J.}~\bibnamefont {Splettstoesser}},\ }\href
  {\doibase 10.1103/PhysRevB.87.245423} {\bibfield  {journal} {\bibinfo
  {journal} {Phys. Rev. B}\ }\textbf {\bibinfo {volume} {87}},\ \bibinfo
  {pages} {245423} (\bibinfo {year} {2013})}\BibitemShut {NoStop}%
\bibitem [{\citenamefont {Battista}\ \emph {et~al.}(2013)\citenamefont
  {Battista}, \citenamefont {Moskalets}, \citenamefont {Albert},\ and\
  \citenamefont {Samuelsson}}]{Battista13}%
  \BibitemOpen
  \bibfield  {author} {\bibinfo {author} {\bibfnamefont {F.}~\bibnamefont
  {Battista}}, \bibinfo {author} {\bibfnamefont {M.}~\bibnamefont {Moskalets}},
  \bibinfo {author} {\bibfnamefont {M.}~\bibnamefont {Albert}}, \ and\ \bibinfo
  {author} {\bibfnamefont {P.}~\bibnamefont {Samuelsson}},\ }\href {\doibase
  10.1103/PhysRevLett.110.126602} {\bibfield  {journal} {\bibinfo  {journal}
  {Phys. Rev. Lett.}\ }\textbf {\bibinfo {volume} {110}},\ \bibinfo {pages}
  {126602} (\bibinfo {year} {2013})}\BibitemShut {NoStop}%
\bibitem [{\citenamefont {Ludovico}\ \emph {et~al.}(2014)\citenamefont
  {Ludovico}, \citenamefont {Lim}, \citenamefont {Moskalets}, \citenamefont
  {Arrachea},\ and\ \citenamefont {S\'anchez}}]{Ludovico14}%
  \BibitemOpen
  \bibfield  {author} {\bibinfo {author} {\bibfnamefont {M.~F.}\ \bibnamefont
  {Ludovico}}, \bibinfo {author} {\bibfnamefont {J.~S.}\ \bibnamefont {Lim}},
  \bibinfo {author} {\bibfnamefont {M.}~\bibnamefont {Moskalets}}, \bibinfo
  {author} {\bibfnamefont {L.}~\bibnamefont {Arrachea}}, \ and\ \bibinfo
  {author} {\bibfnamefont {D.}~\bibnamefont {S\'anchez}},\ }\href {\doibase
  10.1103/PhysRevB.89.161306} {\bibfield  {journal} {\bibinfo  {journal} {Phys.
  Rev. B}\ }\textbf {\bibinfo {volume} {89}},\ \bibinfo {pages} {161306}
  (\bibinfo {year} {2014})}\BibitemShut {NoStop}%
\bibitem [{\citenamefont {Zhou}\ \emph {et~al.}(2015)\citenamefont {Zhou},
  \citenamefont {Thingna}, \citenamefont {H\"anggi}, \citenamefont {Wang},\
  and\ \citenamefont {Li}}]{Zhou15}%
  \BibitemOpen
  \bibfield  {author} {\bibinfo {author} {\bibfnamefont {H.}~\bibnamefont
  {Zhou}}, \bibinfo {author} {\bibfnamefont {J.}~\bibnamefont {Thingna}},
  \bibinfo {author} {\bibfnamefont {P.}~\bibnamefont {H\"anggi}}, \bibinfo
  {author} {\bibfnamefont {J.-S.}\ \bibnamefont {Wang}}, \ and\ \bibinfo
  {author} {\bibfnamefont {B.}~\bibnamefont {Li}},\ }\href {\doibase
  10.1038/srep14870} {\bibfield  {journal} {\bibinfo  {journal} {Scientific
  Reports}\ }\textbf {\bibinfo {volume} {5}},\ \bibinfo {pages} {14870}
  (\bibinfo {year} {2015})}\BibitemShut {NoStop}%
\bibitem [{\citenamefont {Fletcher}\ \emph {et~al.}(2013)\citenamefont
  {Fletcher}, \citenamefont {See}, \citenamefont {Howe}, \citenamefont
  {Pepper}, \citenamefont {Giblin}, \citenamefont {Griffiths}, \citenamefont
  {Jones}, \citenamefont {Farrer}, \citenamefont {Ritchie}, \citenamefont
  {Janssen},\ and\ \citenamefont {Kataoka}}]{Fletcher13}%
  \BibitemOpen
  \bibfield  {author} {\bibinfo {author} {\bibfnamefont {J.~D.}\ \bibnamefont
  {Fletcher}}, \bibinfo {author} {\bibfnamefont {P.}~\bibnamefont {See}},
  \bibinfo {author} {\bibfnamefont {H.}~\bibnamefont {Howe}}, \bibinfo {author}
  {\bibfnamefont {M.}~\bibnamefont {Pepper}}, \bibinfo {author} {\bibfnamefont
  {S.~P.}\ \bibnamefont {Giblin}}, \bibinfo {author} {\bibfnamefont {J.~P.}\
  \bibnamefont {Griffiths}}, \bibinfo {author} {\bibfnamefont {G.~A.~C.}\
  \bibnamefont {Jones}}, \bibinfo {author} {\bibfnamefont {I.}~\bibnamefont
  {Farrer}}, \bibinfo {author} {\bibfnamefont {D.~A.}\ \bibnamefont {Ritchie}},
  \bibinfo {author} {\bibfnamefont {T.~J. B.~M.}\ \bibnamefont {Janssen}}, \
  and\ \bibinfo {author} {\bibfnamefont {M.}~\bibnamefont {Kataoka}},\ }\href
  {\doibase 10.1103/PhysRevLett.111.216807} {\bibfield  {journal} {\bibinfo
  {journal} {Phys. Rev. Lett.}\ }\textbf {\bibinfo {volume} {111}},\ \bibinfo
  {pages} {216807} (\bibinfo {year} {2013})}\BibitemShut {NoStop}%
\bibitem [{\citenamefont {Ubbelohde}\ \emph {et~al.}(2015)\citenamefont
  {Ubbelohde}, \citenamefont {Hohls}, \citenamefont {Kashcheyevs},
  \citenamefont {Wagner}, \citenamefont {Fricke}, \citenamefont {K{\"a}stner},
  \citenamefont {Pierz}, \citenamefont {Schumacher},\ and\ \citenamefont
  {Haug}}]{Ubbelohde15}%
  \BibitemOpen
  \bibfield  {author} {\bibinfo {author} {\bibfnamefont {N.}~\bibnamefont
  {Ubbelohde}}, \bibinfo {author} {\bibfnamefont {F.}~\bibnamefont {Hohls}},
  \bibinfo {author} {\bibfnamefont {V.}~\bibnamefont {Kashcheyevs}}, \bibinfo
  {author} {\bibfnamefont {T.}~\bibnamefont {Wagner}}, \bibinfo {author}
  {\bibfnamefont {L.}~\bibnamefont {Fricke}}, \bibinfo {author} {\bibfnamefont
  {B.}~\bibnamefont {K{\"a}stner}}, \bibinfo {author} {\bibfnamefont
  {K.}~\bibnamefont {Pierz}}, \bibinfo {author} {\bibfnamefont {H.~W.}\
  \bibnamefont {Schumacher}}, \ and\ \bibinfo {author} {\bibfnamefont {R.~J.}\
  \bibnamefont {Haug}},\ }\href {http://dx.doi.org/10.1038/nnano.2014.275}
  {\bibfield  {journal} {\bibinfo  {journal} {Nat. Nanotechnol.}\ }\textbf
  {\bibinfo {volume} {10}},\ \bibinfo {pages} {46} (\bibinfo {year}
  {2015})}\BibitemShut {NoStop}%
\bibitem [{\citenamefont {Gasparinetti}\ \emph {et~al.}(2015)\citenamefont
  {Gasparinetti}, \citenamefont {Viisanen}, \citenamefont {Saira},
  \citenamefont {Faivre}, \citenamefont {Arzeo}, \citenamefont {Meschke},\ and\
  \citenamefont {Pekola}}]{Gasparinetti15}%
  \BibitemOpen
  \bibfield  {author} {\bibinfo {author} {\bibfnamefont {S.}~\bibnamefont
  {Gasparinetti}}, \bibinfo {author} {\bibfnamefont {K.~L.}\ \bibnamefont
  {Viisanen}}, \bibinfo {author} {\bibfnamefont {O.-P.}\ \bibnamefont {Saira}},
  \bibinfo {author} {\bibfnamefont {T.}~\bibnamefont {Faivre}}, \bibinfo
  {author} {\bibfnamefont {M.}~\bibnamefont {Arzeo}}, \bibinfo {author}
  {\bibfnamefont {M.}~\bibnamefont {Meschke}}, \ and\ \bibinfo {author}
  {\bibfnamefont {J.~P.}\ \bibnamefont {Pekola}},\ }\href {\doibase
  10.1103/PhysRevApplied.3.014007} {\bibfield  {journal} {\bibinfo  {journal}
  {Phys. Rev. Applied}\ }\textbf {\bibinfo {volume} {3}},\ \bibinfo {pages}
  {014007} (\bibinfo {year} {2015})}\BibitemShut {NoStop}%
\bibitem [{\citenamefont {B\"uttiker}\ \emph {et~al.}(1993)\citenamefont
  {B\"uttiker}, \citenamefont {Pr\^etre},\ and\ \citenamefont
  {Thomas}}]{Buettiker93}%
  \BibitemOpen
  \bibfield  {author} {\bibinfo {author} {\bibfnamefont {M.}~\bibnamefont
  {B\"uttiker}}, \bibinfo {author} {\bibfnamefont {A.}~\bibnamefont
  {Pr\^etre}}, \ and\ \bibinfo {author} {\bibfnamefont {H.}~\bibnamefont
  {Thomas}},\ }\href {\doibase 10.1103/PhysRevLett.70.4114} {\bibfield
  {journal} {\bibinfo  {journal} {Phys. Rev. Lett.}\ }\textbf {\bibinfo
  {volume} {70}},\ \bibinfo {pages} {4114} (\bibinfo {year}
  {1993})}\BibitemShut {NoStop}%
\bibitem [{\citenamefont {B\"uttiker}\ \emph {et~al.}(1994)\citenamefont
  {B\"uttiker}, \citenamefont {Thomas},\ and\ \citenamefont
  {Pr\^etre}}]{Buettiker94}%
  \BibitemOpen
  \bibfield  {author} {\bibinfo {author} {\bibfnamefont {M.}~\bibnamefont
  {B\"uttiker}}, \bibinfo {author} {\bibfnamefont {H.}~\bibnamefont {Thomas}},
  \ and\ \bibinfo {author} {\bibfnamefont {A.}~\bibnamefont {Pr\^etre}},\
  }\href {\doibase 10.1007/BF01307664} {\bibfield  {journal} {\bibinfo
  {journal} {Z. Phys. B Con. Mat.}\ }\textbf {\bibinfo {volume} {94}},\
  \bibinfo {pages} {133} (\bibinfo {year} {1994})}\BibitemShut {NoStop}%
\bibitem [{\citenamefont {Splettstoesser}\ \emph {et~al.}(2010)\citenamefont
  {Splettstoesser}, \citenamefont {Governale}, \citenamefont {K\"onig},\ and\
  \citenamefont {B\"uttiker}}]{Splettstoesser10}%
  \BibitemOpen
  \bibfield  {author} {\bibinfo {author} {\bibfnamefont {J.}~\bibnamefont
  {Splettstoesser}}, \bibinfo {author} {\bibfnamefont {M.}~\bibnamefont
  {Governale}}, \bibinfo {author} {\bibfnamefont {J.}~\bibnamefont {K\"onig}},
  \ and\ \bibinfo {author} {\bibfnamefont {M.}~\bibnamefont {B\"uttiker}},\
  }\href {\doibase 10.1103/PhysRevB.81.165318} {\bibfield  {journal} {\bibinfo
  {journal} {Phys. Rev. B}\ }\textbf {\bibinfo {volume} {81}},\ \bibinfo
  {pages} {165318} (\bibinfo {year} {2010})}\BibitemShut {NoStop}%
\bibitem [{\citenamefont {Contreras-Pulido}\ \emph {et~al.}(2012)\citenamefont
  {Contreras-Pulido}, \citenamefont {Splettstoesser}, \citenamefont
  {Governale}, \citenamefont {K\"onig},\ and\ \citenamefont
  {B\"uttiker}}]{Contreras12}%
  \BibitemOpen
  \bibfield  {author} {\bibinfo {author} {\bibfnamefont {L.~D.}\ \bibnamefont
  {Contreras-Pulido}}, \bibinfo {author} {\bibfnamefont {J.}~\bibnamefont
  {Splettstoesser}}, \bibinfo {author} {\bibfnamefont {M.}~\bibnamefont
  {Governale}}, \bibinfo {author} {\bibfnamefont {J.}~\bibnamefont {K\"onig}},
  \ and\ \bibinfo {author} {\bibfnamefont {M.}~\bibnamefont {B\"uttiker}},\
  }\href {\doibase 10.1103/PhysRevB.85.075301} {\bibfield  {journal} {\bibinfo
  {journal} {Phys. Rev. B}\ }\textbf {\bibinfo {volume} {85}},\ \bibinfo
  {pages} {075301} (\bibinfo {year} {2012})}\BibitemShut {NoStop}%
\bibitem [{\citenamefont {Saptsov}\ and\ \citenamefont
  {Wegewijs}(2012)}]{Saptsov12a}%
  \BibitemOpen
  \bibfield  {author} {\bibinfo {author} {\bibfnamefont {R.~B.}\ \bibnamefont
  {Saptsov}}\ and\ \bibinfo {author} {\bibfnamefont {M.~R.}\ \bibnamefont
  {Wegewijs}},\ }\href {\doibase 10.1103/PhysRevB.86.235432} {\bibfield
  {journal} {\bibinfo  {journal} {Phys. Rev. B}\ }\textbf {\bibinfo {volume}
  {86}},\ \bibinfo {pages} {235432} (\bibinfo {year} {2012})}\BibitemShut
  {NoStop}%
\bibitem [{\citenamefont {Saptsov}\ and\ \citenamefont
  {Wegewijs}(2014)}]{Saptsov14a}%
  \BibitemOpen
  \bibfield  {author} {\bibinfo {author} {\bibfnamefont {R.~B.}\ \bibnamefont
  {Saptsov}}\ and\ \bibinfo {author} {\bibfnamefont {M.~R.}\ \bibnamefont
  {Wegewijs}},\ }\href {\doibase 10.1103/PhysRevB.90.045407} {\bibfield
  {journal} {\bibinfo  {journal} {Phys. Rev. B}\ }\textbf {\bibinfo {volume}
  {90}},\ \bibinfo {pages} {045407} (\bibinfo {year} {2014})}\BibitemShut
  {NoStop}%
\bibitem [{\citenamefont {Schulenborg}\ \emph {et~al.}(2014)\citenamefont
  {Schulenborg}, \citenamefont {Splettstoesser}, \citenamefont {Governale},\
  and\ \citenamefont {Contreras-Pulido}}]{Schulenborg14a}%
  \BibitemOpen
  \bibfield  {author} {\bibinfo {author} {\bibfnamefont {J.}~\bibnamefont
  {Schulenborg}}, \bibinfo {author} {\bibfnamefont {J.}~\bibnamefont
  {Splettstoesser}}, \bibinfo {author} {\bibfnamefont {M.}~\bibnamefont
  {Governale}}, \ and\ \bibinfo {author} {\bibfnamefont {L.~D.}\ \bibnamefont
  {Contreras-Pulido}},\ }\href {\doibase 10.1103/PhysRevB.89.195305} {\bibfield
   {journal} {\bibinfo  {journal} {Phys. Rev. B}\ }\textbf {\bibinfo {volume}
  {89}},\ \bibinfo {pages} {195305} (\bibinfo {year} {2014})}\BibitemShut
  {NoStop}%
\bibitem [{\citenamefont {Anderson}(1975)}]{Anderson75}%
  \BibitemOpen
  \bibfield  {author} {\bibinfo {author} {\bibfnamefont {P.~W.}\ \bibnamefont
  {Anderson}},\ }\href {\doibase 10.1103/PhysRevLett.34.953} {\bibfield
  {journal} {\bibinfo  {journal} {Phys. Rev. Lett.}\ }\textbf {\bibinfo
  {volume} {34}},\ \bibinfo {pages} {953} (\bibinfo {year} {1975})}\BibitemShut
  {NoStop}%
\bibitem [{\citenamefont {Wick}\ \emph {et~al.}(1952)\citenamefont {Wick},
  \citenamefont {Wightman},\ and\ \citenamefont {Wigner}}]{Wick52}%
  \BibitemOpen
  \bibfield  {author} {\bibinfo {author} {\bibfnamefont {G.~C.}\ \bibnamefont
  {Wick}}, \bibinfo {author} {\bibfnamefont {A.~S.}\ \bibnamefont {Wightman}},
  \ and\ \bibinfo {author} {\bibfnamefont {E.~P.}\ \bibnamefont {Wigner}},\
  }\href {\doibase 10.1103/PhysRev.88.101} {\bibfield  {journal} {\bibinfo
  {journal} {Phys. Rev.}\ }\textbf {\bibinfo {volume} {88}},\ \bibinfo {pages}
  {101} (\bibinfo {year} {1952})}\BibitemShut {NoStop}%
\bibitem [{\citenamefont {Aharonov}\ and\ \citenamefont
  {Susskind}(1967)}]{Aharonov67}%
  \BibitemOpen
  \bibfield  {author} {\bibinfo {author} {\bibfnamefont {Y.}~\bibnamefont
  {Aharonov}}\ and\ \bibinfo {author} {\bibfnamefont {L.}~\bibnamefont
  {Susskind}},\ }\href {\doibase 10.1103/PhysRev.155.1428} {\bibfield
  {journal} {\bibinfo  {journal} {Phys. Rev.}\ }\textbf {\bibinfo {volume}
  {155}},\ \bibinfo {pages} {1428} (\bibinfo {year} {1967})}\BibitemShut
  {NoStop}%
\bibitem [{\citenamefont {Streater}\ and\ \citenamefont
  {Wightman}(2000)}]{Streaterbook}%
  \BibitemOpen
  \bibfield  {author} {\bibinfo {author} {\bibfnamefont {R.}~\bibnamefont
  {Streater}}\ and\ \bibinfo {author} {\bibfnamefont {A.}~\bibnamefont
  {Wightman}},\ }\href@noop {} {\emph {\bibinfo {title} {PCT, spin and
  statistics, and all that}}},\ Landmarks in Physics\ (\bibinfo  {publisher}
  {Princeton University Press},\ \bibinfo {year} {2000})\BibitemShut {NoStop}%
\bibitem [{\citenamefont {Bogolubov}\ \emph {et~al.}(1987)\citenamefont
  {Bogolubov}, \citenamefont {Logunov}, \citenamefont {Oksak},\ and\
  \citenamefont {Todorov}}]{Bogolubov89}%
  \BibitemOpen
  \bibfield  {author} {\bibinfo {author} {\bibfnamefont {N.~N.}\ \bibnamefont
  {Bogolubov}}, \bibinfo {author} {\bibfnamefont {A.~A.}\ \bibnamefont
  {Logunov}}, \bibinfo {author} {\bibfnamefont {A.~I.}\ \bibnamefont {Oksak}},
  \ and\ \bibinfo {author} {\bibfnamefont {I.~T.}\ \bibnamefont {Todorov}},\
  }\href@noop {} {\emph {\bibinfo {title} {General principles of quantum field
  theory}}}\ (\bibinfo  {publisher} {Nauka},\ \bibinfo {address} {Moscow},\
  \bibinfo {year} {1987})\ \bibinfo {note} {in Russian. English translation:
  Kluwer Academic Publishers, Dordrecht, 1989}\BibitemShut {NoStop}%
\bibitem [{\citenamefont {Timm}(2008)}]{Timm08}%
  \BibitemOpen
  \bibfield  {author} {\bibinfo {author} {\bibfnamefont {C.}~\bibnamefont
  {Timm}},\ }\href {\doibase 10.1103/PhysRevB.77.195416} {\bibfield  {journal}
  {\bibinfo  {journal} {Phys. Rev. B}\ }\textbf {\bibinfo {volume} {77}},\
  \bibinfo {pages} {195416} (\bibinfo {year} {2008})}\BibitemShut {NoStop}%
\bibitem [{\citenamefont {Cohen}\ \emph {et~al.}(2013)\citenamefont {Cohen},
  \citenamefont {Gull}, \citenamefont {Reichman}, \citenamefont {Millis},\ and\
  \citenamefont {Rabani}}]{Cohen13a}%
  \BibitemOpen
  \bibfield  {author} {\bibinfo {author} {\bibfnamefont {G.}~\bibnamefont
  {Cohen}}, \bibinfo {author} {\bibfnamefont {E.}~\bibnamefont {Gull}},
  \bibinfo {author} {\bibfnamefont {D.~R.}\ \bibnamefont {Reichman}}, \bibinfo
  {author} {\bibfnamefont {A.~J.}\ \bibnamefont {Millis}}, \ and\ \bibinfo
  {author} {\bibfnamefont {E.}~\bibnamefont {Rabani}},\ }\href {\doibase
  10.1103/PhysRevB.87.195108} {\bibfield  {journal} {\bibinfo  {journal} {Phys.
  Rev. B}\ }\textbf {\bibinfo {volume} {87}},\ \bibinfo {pages} {195108}
  (\bibinfo {year} {2013})}\BibitemShut {NoStop}%
\bibitem [{\citenamefont {Karlewski}\ and\ \citenamefont
  {Marthaler}(2014)}]{Karlewski14}%
  \BibitemOpen
  \bibfield  {author} {\bibinfo {author} {\bibfnamefont {C.}~\bibnamefont
  {Karlewski}}\ and\ \bibinfo {author} {\bibfnamefont {M.}~\bibnamefont
  {Marthaler}},\ }\href {\doibase 10.1103/PhysRevB.90.104302} {\bibfield
  {journal} {\bibinfo  {journal} {Phys. Rev. B}\ }\textbf {\bibinfo {volume}
  {90}},\ \bibinfo {pages} {104302} (\bibinfo {year} {2014})}\BibitemShut
  {NoStop}%
\bibitem [{\citenamefont {Schoeller}\ and\ \citenamefont
  {Reininghaus}(2009)}]{Schoeller09b}%
  \BibitemOpen
  \bibfield  {author} {\bibinfo {author} {\bibfnamefont {H.}~\bibnamefont
  {Schoeller}}\ and\ \bibinfo {author} {\bibfnamefont {F.}~\bibnamefont
  {Reininghaus}},\ }\href {\doibase 10.1103/PhysRevB.80.045117} {\bibfield
  {journal} {\bibinfo  {journal} {Phys. Rev. B}\ }\textbf {\bibinfo {volume}
  {80}},\ \bibinfo {pages} {045117} (\bibinfo {year} {2009})}\BibitemShut
  {NoStop}%
\bibitem [{\citenamefont {Schulenborg}\ \emph {et~al.}()\citenamefont
  {Schulenborg}, \citenamefont {Saptsov}, \citenamefont {Haupt}, \citenamefont
  {Splettstoesser},\ and\ \citenamefont {Wegewijs}}]{Schulenborg15Suppmat}%
  \BibitemOpen
  \bibfield  {author} {\bibinfo {author} {\bibfnamefont {J.}~\bibnamefont
  {Schulenborg}}, \bibinfo {author} {\bibfnamefont {R.~B.}\ \bibnamefont
  {Saptsov}}, \bibinfo {author} {\bibfnamefont {F.}~\bibnamefont {Haupt}},
  \bibinfo {author} {\bibfnamefont {J.}~\bibnamefont {Splettstoesser}}, \ and\
  \bibinfo {author} {\bibfnamefont {M.~R.}\ \bibnamefont {Wegewijs}},\
  }\href@noop {} {}\bibinfo {note} {Supplementary material}\BibitemShut
  {NoStop}%
\bibitem [{\citenamefont {Breuer}\ and\ \citenamefont
  {Petruccione}(2002)}]{Breuer}%
  \BibitemOpen
  \bibfield  {author} {\bibinfo {author} {\bibfnamefont {H.-P.}\ \bibnamefont
  {Breuer}}\ and\ \bibinfo {author} {\bibfnamefont {F.}~\bibnamefont
  {Petruccione}},\ }\href@noop {} {\emph {\bibinfo {title} {The Theory of Open
  Quantum Systems}}}\ (\bibinfo  {publisher} {Oxford University Press},\
  \bibinfo {year} {2002})\BibitemShut {NoStop}%
\bibitem [{\citenamefont {Fujisawa}\ \emph {et~al.}(2003)\citenamefont
  {Fujisawa}, \citenamefont {Austing}, \citenamefont {Tokura}, \citenamefont
  {Hirayama},\ and\ \citenamefont {Tarucha}}]{Fujisawa03rev}%
  \BibitemOpen
  \bibfield  {author} {\bibinfo {author} {\bibfnamefont {T.}~\bibnamefont
  {Fujisawa}}, \bibinfo {author} {\bibfnamefont {D.~G.}\ \bibnamefont
  {Austing}}, \bibinfo {author} {\bibfnamefont {Y.}~\bibnamefont {Tokura}},
  \bibinfo {author} {\bibfnamefont {Y.}~\bibnamefont {Hirayama}}, \ and\
  \bibinfo {author} {\bibfnamefont {S.}~\bibnamefont {Tarucha}},\ }\href
  {http://stacks.iop.org/0953-8984/15/i=33/a=201} {\bibfield  {journal}
  {\bibinfo  {journal} {J. Phys. Cond. Mat.}\ }\textbf {\bibinfo {volume}
  {15}},\ \bibinfo {pages} {R1395} (\bibinfo {year} {2003})}\BibitemShut
  {NoStop}%
\bibitem [{\citenamefont {R\"ossler}\ \emph {et~al.}(2013)\citenamefont
  {R\"ossler}, \citenamefont {Kr\"ahenmann}, \citenamefont {Baer},
  \citenamefont {Ihn}, \citenamefont {Ensslin}, \citenamefont {Reichl},\ and\
  \citenamefont {Wegscheider}}]{Roessler13}%
  \BibitemOpen
  \bibfield  {author} {\bibinfo {author} {\bibfnamefont {C.}~\bibnamefont
  {R\"ossler}}, \bibinfo {author} {\bibfnamefont {T.}~\bibnamefont
  {Kr\"ahenmann}}, \bibinfo {author} {\bibfnamefont {S.}~\bibnamefont {Baer}},
  \bibinfo {author} {\bibfnamefont {T.}~\bibnamefont {Ihn}}, \bibinfo {author}
  {\bibfnamefont {K.}~\bibnamefont {Ensslin}}, \bibinfo {author} {\bibfnamefont
  {C.}~\bibnamefont {Reichl}}, \ and\ \bibinfo {author} {\bibfnamefont
  {W.}~\bibnamefont {Wegscheider}},\ }\href
  {http://stacks.iop.org/1367-2630/15/i=3/a=033011} {\bibfield  {journal}
  {\bibinfo  {journal} {New J. Phys.}\ }\textbf {\bibinfo {volume} {15}},\
  \bibinfo {pages} {033011} (\bibinfo {year} {2013})}\BibitemShut {NoStop}%
\end{thebibliography}
%
\onecolumngrid
\cleardoublepage



\section*{Supplementary material}

J. Schulenborg[(1),](*) R. B. Saptsov[(2,3),](*) F. Haupt[(3,4)], J. Splettstoesser[(1)], and M. R. Wegewijs[(2,3,5)]

*(1) Department of Microtechnology and Nanoscience (MC2),*
*Chalmers University of Technology, SE-41298 Göteborg, Sweden*
*(2) Institute for Theory of Statistical Physics, RWTH Aachen, 52056 Aachen, Germany*
*(3) JARA- Fundamentals of Future Information Technology*
*(4) JARA Institute for Quantum Information, RWTH Aachen, 52056 Aachen, Germany*
*(5) Peter Grünberg Institut, Forschungszentrum Jülich, 52425 Jülich, Germany*

## CONTENTS

The supplementary information is divided into two parts:

1. In Sec. I, we provide some details for the weak-coupling, high-temperature master equation calculations.

2. In Sec. II, we write out the details of the derivation outlined at the end of the main article.

*Within this supplementary material, references are numbered as, e.g., equation (S-1) and Figure S-1, whereas regular numbers, e.g., equation (1) and Figure 1, refer to the main article.*

---

* First and second author contributed equally.

# I. ENERGY DECAY IN A QUANTUM DOT

## A. Master equation [Eq. (2)]

### 1. Golden-rule transition rates

In the main article, the master equation [Eq. (2)] is written in abstract form to enable the fermion-parity duality analysis in its clearest formulation using operators rather than matrix representations. Here we give the explicit form in terms of occupation probabilities, which may be more familiar. To this end, one represents the density operator in the orthogonal basis of physical states $i = 0, 1$ or $2$ electrons, i.e., the positive operators with unit trace $[(\mathbb{1}|i)]$: $|0) := |0\rangle\langle 0|$ (pure), $|1) := \frac{1}{2}\sum_\sigma |\sigma\rangle\langle\sigma|$ (mixed), and $|2) := |2\rangle\langle 2|$ (pure). Inserting a completeness relation $\sum_i |i)(i|/(i|i) = 1$, accounting for the non-unit scalar product $(1|1) = \frac{1}{2}$, the density operator reads $|\rho(t)) = \sum_i |i)P_i$ with $P_i := (i|\rho(t))/(i|i)$ being the physical occupations of state $i = 0, 1$, and 2. Similarly, inserting a completeness relation to the left and right of $W$, we can express the kernel in terms of state transition rates:

$$W = \sum_{ij} W_{ij} |i)(j| \quad , \quad W_{ij} = \frac{(i|W|j)}{(i|i)(j|j)}, \tag{S-1}$$

where $W_{ij}$ is the rate for a transition from state $|j)$ to state $|i)$. Substituted into the abstract equation $\partial_t |\rho(t)) = W|\rho(t))$, we obtain the usual master equation for the probabilities

$$\frac{d}{dt}\begin{bmatrix} P_0(t) \\ P_1(t) \\ P_2(t) \end{bmatrix} = \frac{1}{2}\Gamma \begin{bmatrix} -2f^+(\epsilon) & f^-(\epsilon) & 0 \\ 2f^+(\epsilon) & -f^-(\epsilon) - f^+(\epsilon + U) & 2f^-(\epsilon + U) \\ 0 & f^+(\epsilon + U) & -2f^-(\epsilon + U) \end{bmatrix} \times \begin{bmatrix} P_0(t) \\ P_1(t) \\ P_2(t) \end{bmatrix}. \tag{S-2}$$

The notation for the Fermi functions has been adopted from the main article: $f^\pm(\epsilon) = (e^{\pm(\epsilon-\mu)/T} + 1)^{-1}$. Notably, the matrix elements are given by $W_{ij} \cdot (j|j)$, and hence include the normalization of the state $|j)$. In the second column of the matrix in Eq. (S-2), this yields $1/2$ times the corresponding expansion coefficients $W_{i1}$. Physically, this factor $1/2$ is caused by the spin-degree of freedom which is "integrated out" by working with the mixed basis state $|1) = \frac{1}{2}\sum_\sigma |\sigma\rangle\langle\sigma|$. The point is that the rate for each elementary tunnel process between states of a *fixed spin* is $\frac{1}{2}\Gamma = \Gamma_\sigma$. For example, in the formula $\frac{d}{dt}P_0 = \frac{1}{2}\Gamma[-2f^+(\epsilon)P_0 + f^-(\epsilon)P_1]$, the first contribution reflects that the $i = 0$ electron state can decay in two ways to the $i = 1$ electron state which comprises of *two* spin states; the second term instead expresses that a state with only $i = 1$ electron and a fixed spin has only *one* channel to decay to the $i = 0$ state.

Finally, we note that the coefficients $W_{20} = W_{02} = 0$, since $W$ is evaluated only to leading, linear order in the tunneling strength $\Gamma$. The reason is that $W_{20}$ and $W_{20}$ are associated with *electron-pair* tunneling, and thus become nonzero only in the next-to-leading order ($\Gamma^2$), see Ref. [1].

### 2. Eigenvectors of the kernel

In the main article, we briefly sketch how to use the duality Eq. (4) in order to derive the left and right eigenvectors of the kernel $W$. However, since the main article does not provide explicit expressions for all vectors, in particular for the charge mode $|c)$ and its amplitude covector $(c'|$, we here give a full list of *all* rates $\gamma_x$ (minus the eigenvalues of $W$) and eigenvectors $|x)$, $(x'|$ for $x = z, p$ and $c$.

As stated in the main article, the kernel must conserve probability and has a unique stationary state. This implies the existence of only one zero eigenvalue with its corresponding left and right eigenvector:

$$\gamma_z = 0 \quad , \quad (z'| = (\mathbb{1}|$$
$$|z) = \frac{1}{f^+(\epsilon) + f^-(\epsilon + U)}\left[f^-(\epsilon)f^-(\epsilon + U)|0) + 2f^+(\epsilon)f^-(\epsilon + U)|1) + f^+(\epsilon)f^+(\epsilon + U)|2)\right]. \tag{S-3}$$

Note that the stationary state $|z)$, given by the Boltzmann factor $z = e^{-(H-\mu N)/T}/\operatorname{Tr}\left[e^{-(H-\mu N)/T}\right]$, has been expressed in terms of the Fermi functions, $f^\pm$, and in terms of the above defined basis states for zero, single and double occupation: $|0), |1), |2)$.

Next, we apply the duality (4) to $(z'|$ and $|z)$ in order to obtain the fermion-parity rate $\gamma_p$, its mode, and the amplitude covector:

$$\gamma_p = \Gamma \quad , \quad (p'| = [\mathcal{PI}|z)]^\dagger = ((-\mathbb{1})^N z_i| \quad , \quad |p) = [(z'|\mathcal{IP}]^\dagger = |(-\mathbb{1})^N)$$

$$|z_i) = \mathcal{I}|z) = \frac{1}{f^-(\epsilon) + f^+(\epsilon + U)} \left[ f^+(\epsilon) f^+(\epsilon + U)|0) + 2 f^-(\epsilon) f^+(\epsilon + U)|1) + f^-(\epsilon) f^-(\epsilon + U)|2) \right]. \quad \text{(S-4)}$$

The state $|z_i)$ is the inverted stationary state of the dual model, obtained by the energy inversion operation $\mathcal{I}$ which effects $H \to -H$ ($\epsilon \to -\epsilon$ and $U \to -U$) and $\mu \to -\mu$. In the paper, we simply express it as the energy inverted Boltzmann factor $z_i = e^{+(H-\mu N)/T} / \text{Tr} \left[ e^{+(H-\mu N)/T} \right]$. Here, this energy inversion is reflected by swapping $f^+ \leftrightarrow f^-$ in the stationary state $|z)$, as written in Eq. (S-3). This proves in particular that $|z_i)$ is a *physical* state with non-negative occupation probabilities.

Finally, the charge mode $|c)$ and its covector $(c'|$ follow, up to a constant normalization factor, from biorthogonality to $(z'|$ and $(p'|$ as well as $|z)$ and $|p)$. We find

$$\gamma_c = \frac{\Gamma}{2} \left[ f^+(\epsilon) + f^-(\epsilon + U) \right] \quad , \quad (c'| = (N| - N_z(\mathbb{1}| \quad , \quad |c) = \frac{1}{2}(-\mathbb{1})^N \left[ |N) - N_i|\mathbb{1}) \right].  \quad \text{(S-5)}$$

Since the duality (4) holds for any exisiting mode or covector, the charge mode $|c)$ must necessarily be related to the amplitude vector $(c'|$, and the charge rate must be self-dual. In accordance, we find $\gamma_c = \Gamma - \mathcal{I}\gamma_c$ and $(c'| = 2 [\mathcal{PI}|c)]^\dagger$. The additional factor 2 is the freely chosen normalization factor, which can be motivated as follows: by explicit calculation, one can show that $(A|p) = (A|(-\mathbb{1})^N) = 0$ for any operator $A$ that is not at least *quartic* in the dot field operators $d_\sigma, d_\sigma^\dagger$. This means that the parity mode $|p)$ only enters the time-dependent decay of two-particle quantities, implying that $|c)$ is the only mode that affects the time evolution of single-particle observables. Since the charge number $N$ is the only relevant single-particle observable of our system, the amplitude $(c'|$ must be related to the charge $N$, and thereby motivates the choice of a normalization factor which yields $(c'| = (N| - N_z(\mathbb{1}|$. Biorthonormality then also fixes the normalization of the mode vector $|c)$.

### 3. Why is the fermion-parity rate the highest rate? [Eq. (5)]

In the main article, we mention that for any system that obeys the general duality relation (1), the parity rate given by the lumped sum of coupling constants $\Gamma$ is the largest decay rate in the Born-Markov limit. This deserves some comment. We restrict our discussion[a] to master equations, i.e., problems for which only the diagonal part of the density operator in the energy ($H$) basis needs to be considered together with the corresponding blocks of the kernel $W$. In this case, the duality relation Eq. (4) dictates $\Gamma$ to be the highest rate due to the following argument: for any mode $x$ with rate $\gamma_x$, there must be another mode $y$ with $\gamma_y = \Gamma - \mathcal{I}\gamma_x$. Applying the parameter substitution $\mathcal{I}$ of Eq. (4) of the main article and using that two substitutions cancel ($\mathcal{I}^2 = 1$), we obtain $\mathcal{I}\gamma_y = \Gamma - \gamma_x$. Finally, if we assume that all rates of the system of interest *as well as of its dual system*[b] are non-negative, the rates must indeed obey $0 \leq \gamma_x \leq \Gamma$ for all modes $x$.

To argue that the dual system is expected to have non-negative rates whenever the original system has non-negative rates, and thus also exhibits exponential decay, let us consider the dual transformation in detail:

- Inversion of the local energies, $H \to \bar{H} = -H$: For models with a finite number of orbitals considered here, the many-particle spectrum is finite and thus stays bounded from below after inversion of the spectrum. Thus, no problems with the existence of a stationary state can arise.

- Inversion of the chemical potentials, $\mu \to -\mu$: In the wide-band limit, one effectively assumes empty reservoir states above $\mu$ and filled reservoir states below $\mu$ to extend up to energy $\pm\infty$. Inverting the sign of $\mu$ does not change this situation. Note that we do *not* invert the electron energies, but only the value of $\mu$ up to which the electron states are occupied.

This means that the dual model comprises a finite set of discrete quantum states coupled to a flat, wide-band reservoir described by the same Born-Markov master equation differing only in the physical parameter values. Therefore, it will exhibit exponential decay whenever the original system of interest shows exponential decay, and the parity rate $\gamma_p = \Gamma$ must indeed be the highest rate.

---

[a] For quantum master equations in which coherences play a role, the following discussion can be extended. This is avoided here, since the consistent treatment of the coherences requires a careful discussion.

[b] In the Born-Markov limit, the dual system has the same coupling $H^T$, cf. discussion before Eq. (4).

## B. Heat-current

### 1. Derivation of the transient charge and heat current

In this section, we derive the Born-Markov limit expressions for the time-dependent particle- and heat current. In particular, we explain how to obtain the excitation amplitudes of the heat current as reported in Eqs. (7),(8) of the main article.

First, we employ the conservation of the total particle number $N^{\text{tot}} = N + N^{\text{R}}$ to express the particle current into the reservoir in terms of the current out of the dot:

$$I_N(t) = \partial_t \left\langle N^{\text{R}} \right\rangle(t) = -\partial_t \left\langle N \right\rangle(t) = -\partial_t \operatorname{Tr}\left[N\rho(t)\right] = -(N|\partial_t|\rho(t)) = -(N|W|\rho(t)). \tag{S-6}$$

We insert the formal eigenmode expansion of $|\rho(t))$ given in Eq. (3), using the expressions for the eigenvectors given in Eqs. (S-3),(S-4),(S-5). In particular, we express the action of the operator $N$ on the kernel $W$ as

$$(N|W = (c'|W = -\gamma_c(c'| = -\gamma_c\left[(N| - N_z(\mathbb{1}|\right], \tag{S-7}$$

where we exploited $(\mathbb{1}|W = (z'|W = 0$. With the trace normalization $(\mathbb{1}|\rho_0) = 1$, we find

$$I_N(t) = -(N|W|\rho(t)) = -(c'|W|\rho(t)) = \gamma_c(c'|\rho(t)) = \gamma_c(N_0 - N_z)e^{-\gamma_c t}. \tag{S-8}$$

Here, $N_0(\epsilon_0) = (N|z(\epsilon_0))$ and $N_z(\epsilon) = (N|z(\epsilon))$ are the particle numbers in the initially prepared state $\rho_0$ and in the final stationary state $z(\epsilon)$ a long time after the level shift $\epsilon_0 \to \epsilon$ (at constant interaction strength $U$ and temperature $T$). The initially prepared state is assumed to be the stationary state with respect to the inital level $\rho_0 = z(\epsilon_0)$. The single exponential decay rate is given by the charge rate $\gamma_c = \frac{1}{2}\Gamma[f^+(\epsilon) + f^-(\epsilon + U)]$.

The energy current into the reservoir is in general not conserved, as the tunnel coupling itself can store energy. Therefore, the energy flow $I_E^{\text{dot}}(t) = -\partial_t \langle H \rangle(t)$ out of the dot differs from the flow $I_E^{\text{R}}(t) = \partial_t \langle H^{\text{R}} \rangle$ into the reservoir. Nevertheless, in the Born-Markov limit, explicit calculations show that $I_E^{\text{R}}(t) \approx I_E^{\text{dot}}(t) = I_E(t)$ up to corrections $\propto \Gamma^2$, i.e., it is consistent to treat the energy current as a conserved quantity. Thus we obtain the formula mentioned in the main article:

$$I_Q(t) = \partial_t \langle H^{\text{R}} - \mu N^{\text{R}} \rangle \approx -\partial_t \langle H - \mu N \rangle = -\partial_t(H - \mu N|\rho(t)) = -(H - \mu N|W|\rho(t)) \tag{S-9}$$

To obtain the energy current out of the dot, we first rewrite $(H| = \epsilon(N| + U(2|$ where $(N|\bullet = \operatorname{Tr} N\bullet$ and $(2|\bullet = \operatorname{Tr}|2\rangle\langle 2|\bullet = \langle 2|\bullet|2\rangle$. Then, we use $(2|p) = \langle 2|(-\mathbb{1})^N|2\rangle = 1$ to obtain

$$(N|W = (c'|W \quad , \quad (2|W = (2|p)(p'|W + (2|c)(c'|W = (p'|W + \tfrac{1}{2}(2 - N_i)(c'|W, \tag{S-10}$$

where $N_i(\epsilon) = (N|z_i(\epsilon))$ is the average particle number in the inverted stationary state $|z_i)$ as defined in Eq. (S-4). With these relations, the energy current is found to be

$$
\begin{aligned}
I_E(t) = -\partial_t(H|\rho(t)) = -(H|W|\rho(t)) &= -\left[\epsilon + \tfrac{1}{2}(2 - N_i)U\right](c'|W|\rho(t)) - U(p'|W|\rho(t)) \\
&= \left[\epsilon + \tfrac{1}{2}(2 - N_i)U\right] \cdot I_N(t) + U\Gamma \cdot (z_i|(-\mathbb{1})^N|\rho_0)e^{-\Gamma t},
\end{aligned} \tag{S-11}
$$

where we have identified the charge current $I_N(t) = -(c'|W|\rho(t))$ as given in Eq. (S-8), and again used the formal eigenmode expansion of $|\rho(t))$ Eq. (3) as well as the eigenvectors Eqs. (S-3),(S-4),(S-5). Finally, for the heat current $I_Q(t) = I_E(t) - \mu I_N(t)$, we obtain the double exponential form

$$I_Q(t) = a_c e^{-\gamma_c t} + a_p e^{-\gamma_p t} \quad , \quad a_c = \left[\epsilon - \mu + \tfrac{1}{2}(2 - N_i(\epsilon))U\right] \times (N_0 - N_z) \times \gamma_c \quad , \quad a_p = U\Gamma\left((-\mathbb{1})^N z_i|\rho_0\right). \tag{S-12}$$

Whereas the charge amplitude $a_c$ already has the shape reported in Eq. (7) of the main article, the parity amplitude here has a different form. The advantage of the form given in Eq. (S-12) is that, when analyzing the coefficients in the formula for $|z_i)$ in Eq. (S-4), one realizes that $(-\mathbb{1})^N|z_i) \approx |z_i)$ for $U \gg T$. This indeed confirms our claim from the main article that $a_p \approx (z_i|\rho_0)U\Gamma$ is given by the overlap between the inverted stationary state and the initial state. However, what is left to show is the equivalence to Eq. (8) from the main paper, stating that

$$a_p/(U\Gamma) = \tfrac{1}{2}(N_i - 1)(N_0 - 1) + \tfrac{1}{4}(p_i + p_0), \tag{S-13}$$

with $N_i$, $p_i$ depending on $\epsilon$, and $N_0$ and $p_0$ depending on $\epsilon_0$. This equivalence follows most easily when noting that the three operators $\mathbb{1}$, $N - \mathbb{1}$ and $(-\mathbb{1})^N$ form an orthogonal basis with scalar product norms $(\mathbb{1}|\mathbb{1}) = ((-\mathbb{1})^N|(-\mathbb{1})^N) = 4$

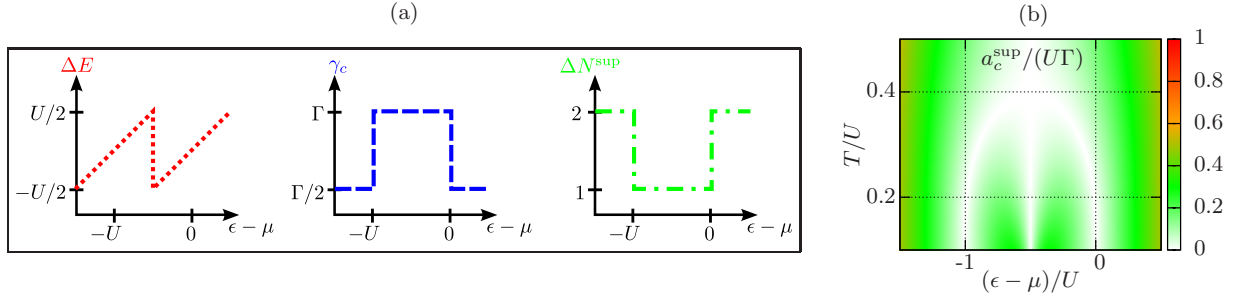

(a)

(b)

FIG. S-1. (a) The three functions entering Eq. (S-19), namely, the energy prefactor $\Delta E$, the charge rate $\gamma_c$ and the upper bound for the excess charge $\Delta N^{\mathrm{sup}}$, as a function of the level position $\epsilon$ with respect to the chemical potential $\mu$ for a fixed interaction strength $U$ in the zero temperature limit $T \to 0$. (b) Upper bound for the charge amplitude, defined by $a_c^{\mathrm{sup}}$ in Eq. (S-19), as a function of the dot level position $\epsilon$ and the temperature $T$. The typical time scale is given by the coupling constant $\Gamma$.

and $(N - \mathbb{1} | N - \mathbb{1}) = 2$. Expanding the initial and *inverted* state in this basis, the coefficients are completely fixed by the average particle numbers $N_0$, $N_i$ and average parities $p_0$, $p_i$ in these states:

$$|\rho_0) = \tfrac{1}{4}|\mathbb{1}) + \tfrac{1}{2}(N_0 - 1)|N - \mathbb{1}) + \tfrac{1}{4}p_0|(-\mathbb{1})^N), \tag{S-14}$$

$$|z_i) = \tfrac{1}{4}|\mathbb{1}) + \tfrac{1}{2}(N_i - 1)|N - \mathbb{1}) + \tfrac{1}{4}p_i|(-\mathbb{1})^N). \tag{S-15}$$

The advantage of this basis is that the action of the parity operator – required in Eq. (S-13) – has the simple effect of swapping the first and last expansion coefficients, since $[(-\mathbb{1})^N]^2 = \mathbb{1}$:

$$(-\mathbb{1})^N|z_i) = \tfrac{1}{4}p_i|\mathbb{1}) + \tfrac{1}{2}(N_i - 1)|N - \mathbb{1}) + \tfrac{1}{4}|(-\mathbb{1})^N). \tag{S-16}$$

Taking the scalar product with $\rho_0$ and accounting for the non-unit norms given above immediately shows that $a_p = U\Gamma\left((-1)^N z_i | \rho_0\right)$ is equivalent to Eq. (S-13).

### 2. Estimate for the charge mode amplitude $a_c$ [Eq. (8) and Fig. 1]

In Fig. 1(d) of the main article, we plotted the charge amplitude $a_c$ for the specific temperature $T/U = 1/10$ and pointed out that $|a_c| \lesssim \tfrac{1}{2}\Gamma U$. Here, we explicitly show that within the plotted window of final level positions $-U < \epsilon - \mu + \tfrac{U}{2} < U$ this holds for all temperatures of immediate experimental interest, $T \ll U$. To discuss this in the most efficient way, we note that according to Eq. (S-12), $|a_c|$ decomposes into three factors:

$$|a_c| = \left| \epsilon - \mu + \tfrac{1}{2}(2 - N_i(\epsilon))U \right| \times \gamma_c \times |N_0 - N_z| . \tag{S-17}$$

The only term that depends on the *initial* level position $\epsilon_0$ is the initial occupation number $N_0$ in the modulus of the excess charge $|N_0 - N_z|$. Hence, to determine an upper bound for $|a_c|$ as a function of *only* the final level $\epsilon$ and the temperature $T$ relative to the scale $U$, we first have to find a supremum for the excess charge $|N_0 - N_z|$ for arbitrary initial levels $\epsilon_0$. Clearly, since the average number of particles on the dot is between 0 and 2, it is always true that

$$0 \leq |N_0 - N_z| \leq \Delta N^{\mathrm{sup}} \equiv \max\left\{ N_z(\epsilon, U, T), 2 - N_z(\epsilon, U, T) \right\}. \tag{S-18}$$

This means that

$$a_c^{\mathrm{sup}} = |\Delta E| \times \gamma_c \times \Delta N^{\mathrm{sup}} \text{ with } \Delta E = \epsilon - \mu + \frac{2 - N_i(\epsilon)}{2}U \tag{S-19}$$

forms an upper bound to the modulus of the charge mode amplitude $|a_c|$. To get an intuition about how $a_c^{\mathrm{sup}}$ behaves as a function of the final level position, Fig. S-1(a) shows the three factors entering Eq. (S-19), that is, the energy prefactor $\Delta E$, the charge rate $\gamma_c$, and the upper bound for the excess charge $\Delta N^{\mathrm{sup}}$, in the limit $T \to 0$. Multiplying the three functions, it becomes intuitively clear that $0 \leq |a_c| \leq a_c^{\mathrm{sup}} \lesssim U\Gamma/2$ inside the Coulomb window $-U < \epsilon - \mu + \tfrac{U}{2} < U$. However, since the limit $T \to 0$ is not strictly allowed to be taken in our weak coupling approach, Fig. S-1(b) also explicitly visualizes $a_c^{\mathrm{sup}}$ as a function of $(\epsilon - \mu)/U$ for temperatures between $T/U = 0.1$ and $T/U = 0.5$. Together, Figs. S-1(a,b) confirm our claim from the main paper that $0 \leq |a_c| \lesssim U\Gamma/2$.

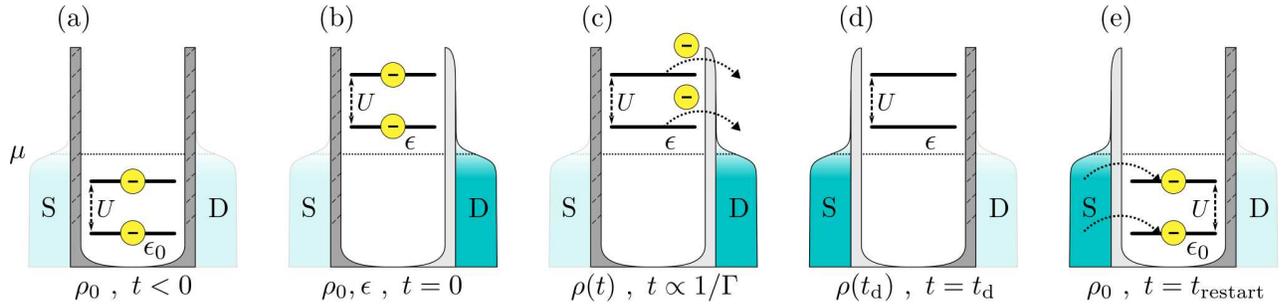

FIG. S-2. Pump-probe scheme described in the text.

### 3. Pump-probe scheme for measuring the time-dependent heat current [Eq. (S-12)]

We briefly sketch a scheme to detect the *time-dependent heat current* [Eq. (S-12)] emitted from the quantum dot by doing a time-averaged temperature measurement in a detector electrode (D). The idea is shown in Fig. S-2 and involves an additional source electrode (S) to refill the dot. It requires independent voltage control over the dot level position and over the heights of the tunnel barriers to S and D, in order to separate the measurement from the refilling process. Furthermore, we assume that the electromagnetic fields arising from the voltage switches themselves do not lead to energy dissipation into the electrodes. Under these conditions, the detailed pump-probe sequence is the following:

(a) At time $t < 0$, the dot is prepared in an initial state $\rho_0$, for example with two electrons as in Fig. S-2. The dot is initially isolated with respect to particle-exchange from both the source electrode (S) and detector electrode (D).

(b) At $t = 0$, a very fast voltage switch (on the time scale $1/\Gamma$) brings the dot into a non-equilibrium situation with respect to the detector electrode by switching the quantum dot energy level $\epsilon_0 \to \epsilon$ (by a primary gate voltage) such that immediately after the switch, the dot is still in its initial state $\rho_0$. Simultaneously, one switches on the tunnel coupling $\Gamma$ to the detector D (by lowering the tunnel barrier height with an appropriate additional gate voltage) to enable the energy measurement.

(c) As discussed in the main article, for $t > 0$ the electron tunneling will induce energy transport into the detector on time scales $1/\gamma_c, 1/\gamma_p \lesssim 1/\Gamma$, described by the heat current Eq. (S-12).

(d) After a delay time $t_d$ the measurement of the heat current is instantly shut off by isolating the dot from the detector (by a voltage switch that raises the tunnel barrier to D) and attaching it to the source S.

(e) The dot energy level is switched back to its initial value $\epsilon \to \epsilon_0$. This starts the preparation of the dot for the next switch. After $t_{\text{restart}} - t_d \gg \Gamma$ the dot has again reached the stationary state $\rho_0$ and the system is ready to return to (a).

Repeating the sequence (a)-(e) of Fig. S-2 a large number of $M$ times, a measurable stationary temperature rise will result in the detector electrode due to $M \times$ the energy that has been deposited in the time interval $t_d$. By measuring this temperature rise as the delay time is varied from $t_d \ll \Gamma^{-1}$ to $\Gamma^{-1} \lesssim t_d$ (but respecting $\Gamma^{-1} \ll t_{\text{restart}} - t_d$), one can extract the full *time-resolved* heat current $I_Q(t)$.

It should be noted that this type of technique requires *fast* voltage control but allows for slow charge and heat measurement, and is well-established in the field of mesoscopic charge detection [2, 3]. Simultaneous measurement of the time-dependent charge and heat current $I_N(t)$ and $I_Q(t)$ would allow a detailed experimental analysis of the time-resolved effects of Coulomb interaction in the heat current as predicted in the main article.

### 4. Extracting the parity rate and the heat current amplitudes [Eq. (5), Eq. (7), Eq. (8)]

In the main article, we claim that a measurement of both the transient charge and heat current after a level shift $\epsilon_0 \to \epsilon$ allows to extract the parity rate $\gamma_p$ and the parity amplitude $a_p$, thereby verifying $\gamma_p = \Gamma$ as well as the signature of attractive interactions in the amplitude in Fig. 1(e). Here, we discuss how this is possible, and also point out some considerations for such experiments.

We first outline the main idea. Using the form we obtained for the time-dependent charge current, $I_N(t) \propto e^{-\gamma_c t}$, and heat current, $I_Q(t) = a_c e^{-\gamma_c t} + a_p e^{-\gamma_p t}$, we can write

$$I_Q(t) = E(\infty) \cdot I_N(t) + a_p e^{-\gamma_p t},$$ (S-20)

where $E(\infty) \propto a_c$ denotes a constant with dimension energy which is to be determined experimentally.

1. If $I_Q(t)$ and $I_N(t)$ are given, from a plot of their ratio one can extract

$$E(\infty) = \frac{I_Q(\infty)}{I_N(\infty)}.$$ (S-21)

Having obtained $E(\infty) \cdot I_N(t)$, the "tight coupling" part of the heat current, one can then construct

$$\delta I_Q(t) := I_Q(t) - E(\infty) \cdot I_N(t) = a_p e^{-\gamma_p t},$$ (S-22)

whose value at $t = 0$ gives $a_p$. In principle, repeating this for various switches $\epsilon_0 \to \epsilon$, one can, without fitting, completely map out the amplitude $a_p$ as shown in Fig. 1(e) of the main article. This would then demonstrate the signature of attractive interaction in the decay of the repulsive system.

2. Next, plotting the normalized time-dependent quantity

$$-\frac{1}{t} \ln \frac{\delta I_Q(t)}{\delta I_Q(0)} = \gamma_p = \Gamma,$$ (S-23)

all data points should collapse onto a single, constant value, independently of which switch $\epsilon_0 \to \epsilon$, which temperature $T$ and which interaction strength $U$ is considered. This would then demonstrate the exceptional restriction of the decay rate $\gamma_p$ enforced by probability conservation, via the duality Eq. (1) based on the fermion-parity superselection principle.

Our results provide an additional, independent check on the consistency of the description: since $a_p = U\Gamma$ whenever it is nonzero [red regions in Fig. 1(e)], dividing the values of $a_p$ by the extracted value for $\gamma_p$ always gives the constant Coulomb energy:

$$a_p/\gamma_p = U.$$ (S-24)

The constancy of this ratio for *all* switches $\epsilon_0 \to \epsilon$ well within the red regions in Fig. 1(e) is again remarkable. This value can be compared with the values of $U$ determined independently by standard *stationary* Coulomb blockade linear-response spectroscopy.

Relevant to the above discussion is the more general, time-dependent ratio with dimension of energy:

$$E(t) = \frac{I_Q(t)}{I_N(t)} = \frac{a_c}{\gamma_c (N_0 - N_z)} + \frac{a_p}{\gamma_c (N_0 - N_z)} e^{-(\gamma_p - \gamma_c)t}.$$ (S-25)

From the expression on the right hand side, obtained by inserting Eq. (S-8) and Eq. (S-12), follow a few experimental considerations for the implementation of the sketched procedure:

- First, the energy levels $\epsilon_0$ and $\epsilon$ should correspond to different charge states of the dot, such that $|N_0(\epsilon_0) - N_z(\epsilon)| \geq 1$, i.e. to ensure that the amplitude of $I_N$ is measurable.

- Second, one should ensure that the time scale set by the difference between the two decay rates, $\gamma_p - \gamma_c$, lies within the window of experimentally accessible times, e.g., of the pump-probe scheme suggested in Sec. I B 3. Interestingly, our duality provides *physical* insight into this rate difference [a] and one finds that the timescale is minimal at $\epsilon - \mu = -U/2$ where the rate-difference scales as $\sim \Gamma e^{-U/(2T)}$, but corrections from higher-order tunneling processes should be expected. To avoid the degeneracy of rates for switches that extend into the Coulomb blockade regime – which expose the attractive-interaction features of the dual model – the temperature $T$ should thus not be too low compared to $U$.

---

[a] This rate difference is always positive since $\gamma_p > \gamma_c$, see Sec. I A 3. In fact, this time scale is set by the charge relaxation rate of the *attractive* dual model: by our duality (4), $\Gamma - \gamma_c(\epsilon, U) = \mathcal{I}\gamma_c(\epsilon, U)$, which equals $\gamma_c(-\epsilon, -U) = \frac{\Gamma}{2} \left[ f^-(\epsilon) + f^+(\epsilon + U) \right]$. It thus makes physical sense that the latter effective rate is minimal at $\epsilon - \mu = -U/2$, where it scales as $\sim \Gamma e^{-U/(2T)}$: here, the dual model exhibits the electron-pair (!) charge degeneracy $N_i = 0 \leftrightarrow 2$ induced by the attractive interaction, and thus allows both an empty and doubly occupied initial state to remain stable.

- Importantly, the difference in amplitudes between the two terms in Eq. (S-25), required for a clear signal, presents no problem: as mentioned in the paper and shown in Sec. I B 2, $a_p \geq 2a_c$, meaning that the amplitude of the time-dependent second term is always larger.

Finally, we note that equation (S-25) also indicates how the charge amplitude $a_c$ in the heat current can be extracted:

$$a_c \approx E(\infty) \cdot \gamma_c \cdot (N_0 - N_z),$$ (S-26)

where the charge difference $N_0 - N_z = \int_0^\infty dt I_N(t)$ is, of course, available, and the required charge rate $\gamma_c$ has also been shown to be experimentally extractable in interacting systems [4].

## II. GENERAL DUALITY IN FERMIONIC OPEN QUANTUM SYSTEMS

The key result of the main paper is the duality relation (1) which is straightforwardly derived in Sec. II B using a formulation of open-system dynamics that was previously established in Refs. [5, 6]. However, to keep the supplement self-contained, Sec. II A first introduces the required background using unified, adapted notation for the reader's convenience.

### A. Previous results and notation

The duality Eq. (1) of the main article concerns the time-evolution of the reduced density operator, $\rho(t) = \Pi(t)\rho(0)$, where $\rho(0)$ is an arbitrary initial state and $\Pi(t)$ the propagator. As done in the vast majority of open-system approaches, we consider reduced dynamics starting from an initially factorized state of the total system $\rho^{\text{tot}}(0) = \rho^{\text{R}}\rho(0)$, where the density operator of the reservoirs $\rho^{\text{R}}$ is described by the grand-canonical ensemble. Then, in general, the propagator is obtained by taking the trace $\text{Tr}_{\text{R}}$ over the fermionic reservoirs of the total system propagator, the solution of the Liouville - von Neumann equation $\frac{d}{dt}\rho^{\text{tot}} = -iL^{\text{tot}}\rho^{\text{tot}}$ with $L^{\text{tot}} = L + L^{\text{R}} + L^{\text{T}}$:

$$\Pi(t) = \text{Tr}_{\text{R}} e^{-iL^{\text{tot}}t} \rho^{\text{R}} = (\mathbb{1}^{\text{R}}|e^{-i(L+L^{\text{R}}+L^{\text{T}})t}|\rho^{\text{R}}).$$ (S-27)

The Liouvillians are the commutators with the respective Hamiltonians, $L^x = [H^x, \bullet]$ for $x = \text{tot}, \text{R}$ or "empty". Here – and in most expressions below – we use the same bra-ket notation [7] for Hilbert-Schmidt scalar products of the reservoir as used in the main article for the subsystem (quantum dot). In this case, for a superoperator $S$ acting on the total system, taking the *partial* matrix element with respect to the reservoirs produces a new superoperator $S'$ that acts only on the subsystem:

$$S' = (\mathbb{1}^{\text{R}}|S|\rho^{\text{R}}) := \text{Tr}_{\text{R}} S\rho^{\text{R}}.$$ (S-28)

This notation has the advantage of clearly distinguishing superoperators from the operators on which they act.

The duality relation (1) can be expressed most conveniently by considering the Laplace-frequency representation $\rho(\omega) = \int_0^\infty dt e^{i\omega t} \rho(t)$ of the time evolution for the reduced density operator $\rho(t) = \Pi(t)\rho(0)$. Using that this operator obeys the quantum kinetic equation

$$\frac{d}{dt}\rho(t) = -iL\rho(t) + \int_0^\infty dt' \mathcal{W}(t-t')\rho(t'),$$ (S-29)

one finds

$$\rho(\omega) = \Pi(\omega)\rho(0) \quad , \quad \Pi(\omega) = \int_0^\infty dt e^{i\omega t} \Pi(t) = \frac{i}{\omega - L - iW(\omega)} \quad , \quad W(\omega) = \int_0^\infty dt e^{i\omega t} \mathcal{W}(t).$$ (S-30)

The aim of the following sections is to derive a duality relation for the propagator $\Pi(\omega)$ that translates to the duality relation (1) for the kernel $W(\omega)$ by applying Eq. (S-30).

#### 1. Reduced dynamics in the wide-band limit

The expansion of the kernel $W$ defined by Eq. (S-30) for the evolution Eq. (S-27) in powers of $L^{\text{T}}$ is the standard perturbation series of the Nakajima-Zwanzig approach [8], which is equivalent [9, 10] to the basic real-time diagrammatic expansion [5, 6, 11, 12]. In the main article, we apply the duality in the very simple Born-Markov limit in which

$W$ is restricted to the leading non-vanishing term in $L^{\mathrm{T}}$:

$$\mathcal{W}(t) \approx -\underset{\mathrm{R}}{\mathrm{Tr}} L^{\mathrm{T}} e^{-i(L^{\mathrm{R}}+L)t} L^{\mathrm{T}} \rho^{\mathrm{R}}. \tag{S-31}$$

For the discussed model, this is equivalent to Fermi's "Golden-Rule" theory. [a] However, the new duality Eq. (1) applies much more generally. The starting point for our derivation is the recent observation that the expansion of $W$ in the coupling Liouvillian can be drastically simplified in the wide band limit *without* assuming weak coupling [5, 6]. Based on earlier renormalization-group ideas [13], this wide-band limit formulation [b] starts from the well-known fact that in the high-temperature limit, the leading order perturbation theory for the exact solution Eq. (S-30) is "good". More precisely, as pointed out in Refs. [5, 6], at infinite reservoir temperature, $T = \infty$, this gives the exact solution in the form [c] $\mathcal{W}(t) = W_\infty \delta(t-0)$, independent of both frequency $\omega$ and the local Liouvillian $L = [H, \bullet]$, and results in strictly Markovian dynamics:

$$\Pi_\infty(t) = \lim_{T \to \infty} \Pi(t) = e^{-i(L+iW_\infty)t} \quad , \quad W_\infty(L^{\mathrm{T}}) = \lim_{T \to \infty} W(\omega; L, L^{\mathrm{T}}) = -\pi \underset{\mathrm{R}}{\mathrm{Tr}} \left[ (L^{\mathrm{T}})^2 \rho_\infty^{\mathrm{R}} \right]. \tag{S-32}$$

Here and below, we only indicate the functional dependence of the kernel $W$ on those parts of $L^{\mathrm{tot}} = L + L^{\mathrm{R}} + L^{\mathrm{T}}$ that are relevant for the discussion of the duality (i.e., the dependence on $L^{\mathrm{R}}$ is not written because it is always assumed to be the same). The key advance made in Refs. [5, 6] was to systematically exploit this result and reorganize the wide-band limit perturbation theory for *any finite* $T$ around this reference solution. This affords many physical and technical insights that remain completely obscured in the original series starting from an uncoupled system ($L^{\mathrm{T}} = 0$). The most fundamental example is the duality (1) reported in the main article, which was not noted yet in Refs. [5, 6].

The main idea of our approach – as sketched in the main article – is that the tunnel coupling is decomposed as $L^{\mathrm{T}} = L_+^{\mathrm{T}} + L_-^{\mathrm{T}}$ into mutually adjoint components $L_\pm^{\mathrm{T}} = (L_\mp^{\mathrm{T}})^\dagger$ as indicated precisely below [Eq. (S-36)]. In Refs. [5, 6], a renormalized perturbation series for $\Pi(\omega)$ was derived which is exact in the wide-band limit: it is generated by expanding the following expression in powers of $L_+^{\mathrm{T}}$ [cf. Eq. (S-61c) below]:

$$\Pi(t) = (\mathbb{1}^{\mathrm{R}} | e^{-i(L^{\mathrm{R}}+L+iW_\infty+L_+^{\mathrm{T}})t} | \rho^{\mathrm{R}}). \tag{S-33}$$

In this much simplified series, one is thus allowed to drop part of the tunnel coupling, $L^{\mathrm{T}} \to L_+^{\mathrm{T}}$, when simply including the $T = \infty$ solution $iW_\infty(L^{\mathrm{T}}) = iW_\infty(L_+^{\mathrm{T}} + L_-^{\mathrm{T}})$ [Eq. (S-32)] into the unperturbed dynamics $L \to L + iW_\infty$. The derivation in Sec. II B of the key result Eq. (1) of the main article boils down to showing that this exact expression for $\Pi(t)$ obeys the duality order-by-order.

### 2. Fermion-parity superselection postulate and open-system dynamics

The mentioned splitting of the coupling Liouvillian $L^{\mathrm{T}}$ into its parts $L_\pm^{\mathrm{T}}$ appears naturally when using special quantum fields [d] which are superoperators [5, 6]. The crucial property that makes these field superoperators such a powerful tool is that they incorporate the structure imposed on the open-system time evolution by the fermion-parity superselection postulate. We now explain at which point this postulate can be exploited.

This superselection postulate requires [14–17] that every physical state $\rho^{\mathrm{tot}}(t)$ and every observable operator commutes at any time with the total fermion-parity operator $(-\mathbb{1})^{N+N^{\mathrm{R}}}$. Thereby, *quantum superpositions* of pure states with even and odd fermion numbers $N^{\mathrm{tot}} = N + N^{\mathrm{R}}$ are prohibited, as mentioned in the introduction of the main article. Equivalently, $|\rho^{\mathrm{tot}}(t))$ must be an eigenvector with eigenvalue $+1$ of the fermion-parity superoperator, which applies the parity operator to both sides of its argument $\bullet$:

$$(-\mathbb{1})^{L^N + L^{N^{\mathrm{R}}}} \bullet := (-\mathbb{1})^{N+N^{\mathrm{R}}} \bullet (-\mathbb{1})^{N+N^{\mathrm{R}}}, \qquad L^N := [N, \bullet]_-, \qquad L^{N^{\mathrm{R}}} := [N^{\mathrm{R}}, \bullet]_-. \tag{S-34}$$

In the Liouville-von Neumann equation $\frac{d}{dt} \rho^{\mathrm{tot}} = -i L^{\mathrm{tot}} \rho^{\mathrm{tot}}$, the Liouvillian $L^{\mathrm{tot}}$ commutes with $(-\mathbb{1})^{L^N + L^{N^{\mathrm{R}}}}$ since the Hamiltonian in $L^{\mathrm{tot}} \bullet = [H^{\mathrm{tot}}, \bullet]$ is dictated to commute with the total fermion-parity operator $(-\mathbb{1})^{N+N^{\mathrm{R}}}$ to

---

[a] Golden-Rule rates are obtained from Eq. (S-31) when restricting it to elements connecting diagonal elements of the density operator (state occupations), which is sufficient for the discussed model. For a general overview of next-to-leading order corrections to the Golden-Rule in various formulations and the complications added by non-diagonal density matrix elements, see Ref. [10].

[b] This is equivalent to the "discrete-step renormalization" in the real-time RG framework of Ref. [12].

[c] The zero in the argument $t-0$ denotes an infinitesimal shift needed when Laplace transforming $\delta$-functions.

[d] These fields are referred to as "causal superfermions" because the algebraic structure of physical expressions in terms of these fields expresses causality requirements.

preserve this requirement for all $t$. Hence, one finds $(-\mathbb{1})^{N+N^{\mathrm{R}}} \bullet (-\mathbb{1})^{N+N^{\mathrm{R}}} = \bullet$ for any argument $\bullet$ appearing in any expression that arises in the *physical* time evolution of the density operator. The splitting $L^{\mathrm{T}} = L_+^{\mathrm{T}} + L_-^{\mathrm{T}}$ appearing in the above wide-band perturbation expansion Eq. (S-33) exploits exactly this consequence of the fermion-parity superselection postulate for fermionic *open quantum system*. Namely, we can replace the coupling Liouvillian $L^{\mathrm{T}}$ by

$$L^{\mathrm{T}} := H^{\mathrm{T}} \bullet - (-\mathbb{1})^{N+N^{\mathrm{R}}} \bullet (-\mathbb{1})^{N+N^{\mathrm{R}}} H^{\mathrm{T}}. \tag{S-35}$$

Thus, all insights that rely essentially on Eq. (S-33) as a starting point ultimately stem from the fermion-parity superselection, in combination with the wide-band limit used to obtain Eq. (S-33). To our knowledge, no other approach makes such explicit use of this principle. Hence, many of the far-reaching consequences of fermion-parity superselection in open-system dynamics have not been noted so far.

### 3. Decomposition of bilinear coupling

Based on the key step Eq. (S-35) enabled by the fermion-parity superselection postulate, the coupling can be split as follows: [a]

$$L^{\mathrm{T}} = L_+^{\mathrm{T}} + L_-^{\mathrm{T}}, \qquad L_q^{\mathrm{T}} = \sum_{21} T_{21} G_2^q J_1^{-q}, \text{ for } q = \pm. \tag{S-36}$$

In Eq. (S-36), $G_2^q$ and $J_1^q$ are special quantum-fields [b] of the open system and reservoir, respectively [c]. They are superoperators, i.e., linear transformations of an operator argument – denoted by $\bullet$ –

$$G_2^q \bullet = \tfrac{1}{\sqrt{2}} \left[ d_2 \bullet + q(-\mathbb{1})^N \bullet (-\mathbb{1})^N d_2 \right], \qquad J_1^q \bullet = \tfrac{1}{\sqrt{2}} \left[ c_1 \bullet + q(-\mathbb{1})^{N^{\mathrm{R}}} \bullet (-\mathbb{1})^{N^{\mathrm{R}}} c_1 \right]. \tag{S-37}$$

The multi-index $1 = \eta_1\sigma_1\omega_1$ labels the reservoir degrees of freedom: $\eta_1 = \pm$ indicates particle / hole, $\sigma_1 = \pm$ corresponds to spin projection $\uparrow, \downarrow$, and $\omega_1$ is the electron energy in the continuous wide band. To make the notation compact, we here use the model system of the main article, namely, a single-level quantum dot attached to a single-band reservoir as an example:

$$c_{\eta\sigma\omega} = c_{\sigma\omega} \text{ for } \eta = - \text{ and } c_{\sigma\omega}^{\dagger} \text{ for } \eta = +, \quad d_{\eta\sigma} = d_{\sigma} \text{ for } \eta = - \text{ and } d_{\sigma}^{\dagger} \text{ for } \eta = +. \tag{S-38}$$

This, however, does not lead to a loss of generality: one can always add additional indices to the field operators denoting additional orbitals, bands etc., without affecting the derivation presented here.

The superoperators (S-37) are proper quantum fields because they obey fermionic anticommutation relations: [d]

$$[G_2^q, G_{2'}^{q'}]_+ = \delta^{q,\bar{q}'}\delta_{\bar{2}2'}\mathbb{1}, \qquad [J_1^q, J_{1'}^{q'}]_+ = \delta^{q,\bar{q}'}\delta_{\bar{1}1'}\mathbb{1}, \tag{S-39}$$

denoting $\bar{q}' = -q'$ for $q' = \pm$. Finally, for the particular model system, the coefficient in Eq. (S-36) reads $T_{21} = \sqrt{\nu_{\sigma_1}} t_{\eta_1\sigma_1}\delta_{\eta_2\bar{\eta}_1}\delta_{\sigma_2\sigma_1}$, which is independent of $\omega_1$ due to the assumption of a wide, flat band with energy-independent density of states $\nu_{\sigma_1}$ and tunnel amplitudes $t_{\eta_1\sigma_1}$ in $H^{\mathrm{T}}$. The latter are defined as $t_{\eta_1\sigma_1} = t_{\sigma_1}$ for $\eta_1 = +$ and $t_{\eta_1\sigma_1} = t_{\sigma_1}^*$ for $\eta_1 = -$. The $\delta$'s express restrictions of particle and spin conservation. This admittedly dense notation is crucial since it captures the essence of a very large class of models to which the duality applies. The generalization of the assumptions and the notation is discussed below in Sec. II A 5.

### 4. Expressing the adjoint in fermion-parity operations

The duality Eq. (1) in the main article involves the adjoints with respect to the Hilbert-Schmidt scalar product between operators $(A|B) = \mathrm{Tr}\, A^{\dagger}B$, i.e., $(A|SB) = (S^{\dagger}A|B)$. Since the evolution (and arbitrary superoperators) can

---

[a] We emphasize that the splitting Eq. (S-36) does *not* correspond to a splitting of the underlying Hamiltonian $H^{\mathrm{T}}$ into two parts, as one verifies explicitly by inserting the definitions Eq. (S-37) to recover (S-35). For this reason, it leads to unexpected new insights in open-system dynamics.

[b] See Ref. [5 and 6] for a physically motivated account of their construction. Note that in contrast to these references, we here do not incorporate the tunnel matrix element $T_{21}$ into the definition of $J_i^q$.

[c] Note that we choose the fields of the reservoir and the local subsystem to commute, $[c, d] = 0$. This is always possible, as shown in Ref. [12].

[d] This shows that the expansion (S-33) is a huge simplification relative to Eq. (S-27): since Eq. (S-33) contains *only one* part of the interaction, $L_+^{\mathrm{T}}$, in which only one type of each field appears for the system (reservoir), namely $G^+$ ($J^-$). Thus, all appearing reservoir fields $J^-$ *anticommute*. This is not the case in Eq. (S-27), in which the additional fields $J^+$ *do not* anticommute with $J^-$ [cf. Eq. (S-39)].

be expressed [6] in the fields (S-37), we need to know their adjoints. It is readily verified from their definition (S-37) that [a]

$$(G_2^q)^\dagger = G_2^{\bar{q}} \quad , \quad (J_1^q)^\dagger = J_1^{\bar{q}}. \tag{S-40}$$

The adjoint thus inverts *two* indices. Namely, $q$ becomes $\bar{q} = -q$, and the multi-indices $\bar{2} = (-\eta_2)\sigma_2$, $\bar{1} = (-\eta_1)\sigma_1\omega_1$ have inverted $\eta_2, \eta_1$ relative to the multi-indices $2 = \eta_2\sigma_2$, $1 = \eta_1\sigma_1\omega_1$, i.e., particle and hole operators are interchanged. [b] An essential step in the derivation of the duality relation (1) is to express the index inversion of the adjoint $(G_2^q)^\dagger$ in Eq. (S-40) using parity operations instead. This can be done by composing three relations established in Ref. [5], and works as follows:

- First, consider the *operator-adjoint transformation* $\mathcal{K}$, which maps an arbitrary operator $\bullet$ into its adjoint

$$\mathcal{K}|\bullet) = |\bullet^\dagger). \tag{S-41}$$

  This operation is of fundamental importance, since $\mathcal{K}\rho = \rho$ must hold for any valid density operator [cf. Eq. (S-47)]. By definition, it inverts the $\eta$ index in the multi-index 2 of the ordinary fields: $\mathcal{K}d_2 = d_{\bar{2}}$. The transformation of the fields $G_2^q$ with $\mathcal{K}$ essentially does the same:

$$\mathcal{K}(G_2^q)\mathcal{K} = qG_2^q(-\mathbb{1})^{L^N}. \tag{S-42}$$

  This follows directly from the definition (S-37), the essence of the derivation [c] being that due to $\mathcal{K}(AB) = B^\dagger A^\dagger$, the operator ordering is reversed in Eq. (S-37), effectively swapping its two terms resulting in a sign $q = \pm$ and a parity operation

$$(-\mathbb{1})^{L^N}\bullet = (-\mathbb{1})^N \bullet (-\mathbb{1})^N. \tag{S-43}$$

  Below, we will use that $(-\mathbb{1})^{L^N}$ anticommutes with all fields $G_2^q$ and thus counts their fermion-parity (see note with Eq. (S-44)) and squares to one, $[(-\mathbb{1})^{L^N}]^2 = 1$.

- Now consider the other, independent parity transformation

$$\mathcal{P}\bullet := (-\mathbb{1})^N \bullet \tag{S-44}$$

  featuring in the central result Eq. (1) of the main article and not to be confused with $(-\mathbb{1})^{L^N} = (-\mathbb{1})^N \bullet (-\mathbb{1})^N$ appearing in Eq. (S-42). The transformation of the fields with $\mathcal{P}$ *only* inverts the $q = \pm$ index:

$$\mathcal{P}(G_2^q)\mathcal{P} = -G_2^{\bar{q}}. \tag{S-45}$$

  This also follows directly [d] from the definition (S-37): essentially, by left-multiplying with $(-\mathbb{1})^N$ and anticommuting it through $d_1\bullet$, the sign of the first term in Eq. (S-37) is inverted, which is equivalent to inverting $q$ and an overall sign.

Composing Eq. (S-40)-(S-45), we see that *physical expressions involving the adjoint* – which are always of the type $[\mathcal{K}L_q^{\mathsf{T}}\mathcal{K}]^\dagger$ (see below) – *can always be expressed in terms of linear fermion-parity operations*:

$$[\mathcal{K}G_2^q\mathcal{K}]^\dagger = q\mathcal{P}(-\mathbb{1})^{L^N}G_2^q\mathcal{P}, \qquad (-\mathbb{1})^{L^N} = (-\mathbb{1})^N \bullet (-\mathbb{1})^N, \qquad \mathcal{P}\bullet = (-\mathbb{1})^N \bullet. \tag{S-46}$$

Equation (S-46) is crucial in proving the duality relation in Sec. II B.

---

[a] See Eq. (S-80) in App. II D 1 and also Eq. (57) from Ref. [5].

[b] One should note that the superscipt index $q = \pm$ in Eq. (S-37) is *not* related to physical real particle / holes: this is rather regulated by the separate index $\eta = \pm$.

[c] See Eq. (S-81) in App. II D 1 below and also Ref. [5] [Eq. (G-6)].

[d] See Eq. (S-82) in App. II D 1 below and also Ref. [5] [Eq. (E-7)-(E-8)].

*5. Applicability to fermionic open-system models*

In the following, we restrict our attention to the simple model of the main article as mentioned with Eq. (S-38). However, the notation is so general that without modification, the derivation of Eq. (1) goes through under much more general conditions: [a]

- *System*: Nowhere did we use special properties of $H$, except that it preserves fermion-parity, as it should by the postulates of quantum mechanics. It describes the local evolution of a finite set of discrete many-fermion quantum states due to *any finite number of orbitals. The Hamiltonian $H$ can be of arbitrary complexity*, including all possible multi-fermion interactions (not just "charging energy" as in the Anderson model). However, we stress that we restrict our attention to purely *fermionic* models $H$, i.e., it does not contain local bosonic degrees of freedom (e.g., as in the Anderson-Holstein model).

- *Reservoirs*: There can be *multiple reservoirs, each with multiple bands*. Extending the multi-index $1 = \eta_1 \sigma_1 \omega_1$ by additional band or reservoir indices takes care of this.

- *Coupling*: We do not have to specify any property of the coupling $H^T$, except that it is bilinear and energy-independent. The special simplification possible in the heat-current problem analysed in the main article, namely, that it conserves spin-projection and total spin length, are thus irrelevant. Each band in each reservoir can have *completely general and independent couplings* to each spin-orbital of the system (within the bilinear, flat, wide-band limit). The only change is that in Eq. (S-54), the last expression on the right for $\Gamma$ will be modified to a different constant by summing over more bare coupling constants. However, it remains an "interface"-constant.

## B. Derivation of the fermion-parity duality relation [Eq. (1)]

We now write out the derivation of the duality relation Eq. (1) in the main article. As mentioned there, it follows as a straightforward consequence of the prior results established in Refs. [5, 6] and reviewed above, even though it requires some notation. The first observation to make is that this duality features the adjoint operation †. This antilinear operation, when applied to physical superoperators, is not tied to any conservation of fundamental properties: for an open system, the dynamics are not unitary, i.e., $\rho(t) \neq U(t)\rho(0)U(t)^\dagger$ [for some unitary $U(t)$], which implies that $\Pi(t)^\dagger \neq \Pi(-t)$. An important ingredient of the derivation is to express the effect of taking the adjoint of the propagator in terms of operations that *do* relate to symmetries that are preserved by open system time evolution. One such fundamental symmetry is the selfadjointness of the reduced density operator $\rho(t)^\dagger = \rho(t)$ that must hold for all times $t$. With $\rho(t) = \Pi(t)\rho(0)$ and the operation $\mathcal{K}$ of taking the operator-adjoint defined in Eq. (S-41), this requirement is

$$\mathcal{K}\Pi(t)\mathcal{K} = \Pi(t). \tag{S-47}$$

This is, of course, satisfied by both the exact general propagator [Eq. (S-27)], but also by the exact wide-band limit propagator (S-33) [and (S-49) below] required here. [b] The derivation of the duality relation Eq. (1) exploits this symmetry by studying the combined *linear operation* $[\mathcal{K} \bullet \mathcal{K}]$ *on* physical *superoperators instead of the antilinear adjoint* †. Just as the adjoint, this operation reverses the order of products:

$$[\mathcal{K}BA\mathcal{K}]^\dagger = [\mathcal{K}A\mathcal{K}]^\dagger[\mathcal{K}B\mathcal{K}]^\dagger. \tag{S-48}$$

By applying $[\mathcal{K} \bullet \mathcal{K}]^\dagger$ to the exact wide-band evolution propagator (S-33),

$$\Pi(t) = (\mathbb{1}^R|e^{-i(L^R + L + iW_\infty + L_+^T)t}|\rho^R), \tag{S-49}$$

we can derive a duality relation for $\Pi(t)$ [see Eq. (S-76)] that, by virtue of the solution to the generalized quantum master equation, Eq. (S-30), is equivalent to the duality relation (1) shown in the main article for the kernel $W$ in the Laplace frequency representation [Sec. II B 3]. The benefit of this procedure is that while the duality is most

---

[a] We note that U(1) symmetry, i.e., particle number conservation is allowed to be broken in the system and by the coupling, but we used particle conservation in the reservoirs [Eq. (S-52)]: $\delta_{21} \propto \delta_{\eta_2, -\eta_1}$. To treat mean-field superconducting electrodes, one may consider lifting this constraint, but the energy-dependent density of states in this case does not allow Eq. (S-33) to be derived from Eq. (S-27). This is an interesting future topic.

[b] One readily verifies $\mathcal{K}\Pi(t)\mathcal{K}$ to be satisfied by the general propagator Eq. (S-27): using $\mathcal{K}iL^x\mathcal{K} = iL^x$ for $x = $ R, T or "empty", one finds $\mathcal{K}\Pi(t)\mathcal{K} = \text{Tr}_R e^{-\mathcal{K}i(L+L^T+L^R)\mathcal{K}t} = \Pi(t)$. In Ref. [5] [Appendix G], the expansion (S-61c) of the exact wide-band limit propagator Eq. (S-33) was explicitly shown to satisfy it order-by-order.

straightforwardly proven in the time-representation, the applications of the duality become more obvious in Laplace frequency space, e.g., if one considers the zero frequency Born-Markov limit as in the main article.

In the following, we proceed in two *physical* stages in the spirit of Refs. [5, 6]: first, we consider the simple infinite temperature $T = \infty$ limit [Sec. II B 1]. Then, we use this to simplify the exact time-evolution for finite temperatures $T < \infty$ [Sec. II B 2].

### 1. Infinite temperature $T = \infty$ limit: exact reference solution

We now first consider the simple solution at $T = \infty$ around which the expansion of (S-49) is organized:

$$\Pi_\infty(t) = e^{-iL_\infty t}, \qquad L_\infty = L + iW_\infty. \tag{S-50}$$

Since $\Pi_\infty$ is a physical limit of a valid time-evolution, it must preserve hermiticity. Indeed, it follows from the definition of the kernel $W_\infty$ in Eq. (S-32) and $L\bullet = [H, \bullet]$ that

$$\mathcal{K}L\mathcal{K} = -L \quad , \quad \mathcal{K}W_\infty\mathcal{K} = W_\infty \quad \Rightarrow \quad \mathcal{K}iL_\infty\mathcal{K} = iL_\infty \quad , \quad \mathcal{K}\Pi_\infty(t)\mathcal{K} = \Pi_\infty(t). \tag{S-51}$$

Below, we use that this also holds for the inverse $\Pi_\infty(t)^{-1} = \Pi_\infty(-t)$ since Eq. (S-51) holds for all $t$, including $t < 0$.

We first derive the duality (1) for this $T = \infty$ case. To this end, we express [a] $W_\infty$, given by Eq. (S-32), in the fields Eq. (S-37):

$$W_\infty = -\sum_{2'1'} \tfrac{1}{2} \Gamma_{2'1'} G_{2'}^+ G_{1'}^-, \qquad \Gamma_{2'1'} := 2\pi \sum_{21} T_{2'2} T_{1'1} \delta_{2\bar{1}}. \tag{S-52}$$

Here and below, the unprimed (primed) multi-indices $1, 2$ ($1', 2'$) refer to the reservoir (local subsystem).

At first sight, no non-trivial duality relation seems to arise in this limit, as the kernel is simply self-adjoint: $W_\infty^\dagger = W_\infty$. However, realizing that selfadjointness does not extend to finite temperatures $T < \infty$, [b] we instead exploit the above discussed symmetry relation $W_\infty^\dagger = [\mathcal{K}W_\infty\mathcal{K}]^\dagger$ to express the adjoint in parity operations via Eq. (S-42), and then find a non-trivial duality that *does* have a finite-temperature extension: by inserting Eq. (S-46) into $W_\infty^\dagger = [\mathcal{K}W_\infty\mathcal{K}]^\dagger$ and using Eq. (S-43), the fields in the expansion (S-52) of $W_\infty$ are permuted [Eq. (S-48)]. To restore their original order, we use the fermionic anticommutation relation, yielding

$$W_\infty(L^T)^\dagger = -\tfrac{1}{2}\mathcal{P}\Big[\sum_{2'1'} \Gamma_{2'1'}(-1)(-\mathbb{1})^{L^N} G_{1'}^-(-\mathbb{1})^{L^N} G_{2'}^+\Big]\mathcal{P} = -\tfrac{1}{2}\mathcal{P}\Big[\sum_{2'1'} \Gamma_{2'1'} G_{1'}^- G_{2'}^+\Big]\mathcal{P}$$

$$= -\tfrac{1}{2}\mathcal{P}\sum_{2'1'} \Gamma_{2'1'}\left(\delta_{2'\bar{1}'} - G_{2'}^+ G_{1'}^-\right)\mathcal{P} = -\Gamma - \mathcal{P}W_\infty(L^T)\mathcal{P}. \tag{S-53}$$

This forces a shift by the real positive number

$$\Gamma = \sum_{2'1'} \tfrac{1}{2}\Gamma_{2'1'}\delta_{2'\bar{1}'} = \sum_{1'} \tfrac{1}{2}\Gamma_{\bar{1}'1'}, \tag{S-54}$$

simplifying to $\Gamma = \sum_\sigma \Gamma_\sigma$ for the simple model of the main article, and an overall sign change. Using that $W_\infty$ is quadratic in $L^T$ according to Eq. (S-32), we can absorb this sign into a dual coupling $\bar{L}^T = iL^T$. With this, we obtain the duality (1) of the main article for the limit $T = \infty$:

$$W_\infty(L^T)^\dagger = -\Gamma + \mathcal{P}W_\infty(\bar{L}^T)\mathcal{P}. \tag{S-55}$$

Its main usefulness for the following lies in the duality that it implies for $\Pi_\infty(t)$: its adjoint can be related to a dual $T = \infty$ evolution $\bar{\Pi}_\infty(t)$:

$$\Pi_\infty(t)^\dagger = [\mathcal{K}\Pi_\infty(t)\mathcal{K}]^\dagger = e^{[-\mathcal{K}i(L+iW_\infty)\mathcal{K}]^\dagger t} = e^{-\Gamma t}\mathcal{P}\bar{\Pi}_\infty(t)\mathcal{P}. \tag{S-56}$$

---

[a] See Eq. (38) and (65) of Ref. [6].

[b] The relation $W_\infty^\dagger = W_\infty$ relies on the fact that both types of fields, $G^+$ and $G^-$, appear in its expression (S-52). Taking the adjoint, one permutes two fermion fields and but also interconverts them, $(G_{1'}^q)^\dagger = G_{\bar{1}}^q$. Using the general property $T_{21} = T_{2\bar{1}}$, the original form is then restored. As a result, left and right eigenvectors of $W_\infty$ are mutually adjoint, as for Hermitian Hamiltonians of closed systems. However, for finite temperature $T < \infty$, no corresponding relation holds because the structure of the finite $T$ corrections is very different: they contain *only* $G^+$ fields, as we will see below [Eq. (S-61c)].

The dual evolution $\bar{\Pi}_\infty(t)$ is generated by a dual frequency-independent dissipative Liouvillian $\bar{L}_\infty$:

$$\bar{\Pi}_\infty(t) = \Pi_\infty(t; \bar{L}, \bar{L}^{\mathrm{T}}) = e^{-i\bar{L}_\infty t}, \qquad \bar{L}_\infty = \bar{L} - iW_\infty(L^{\mathrm{T}}) = \bar{L} + iW_\infty(\bar{L}^{\mathrm{T}}). \tag{S-57}$$

Here and below, dual quantities are indicated by an overbar. [a] They determine the dynamics of an underlying dual model with the same reservoir Liouvillian ($\bar{L}^{\mathrm{R}} = L^{\mathrm{R}}$ is never written), but with inverted energy signs in the local Liouvillian $L$ and couplings multiplied by $i$, recalling a Wick rotation:

$$\bar{L} := -L, \qquad \bar{L}^{\mathrm{T}} := iL^{\mathrm{T}}. \tag{S-58}$$

What remains to be proven in the following is that the duality (S-56) with the dual Liouvillians given in Eq. (S-57) extends to finite temperatures without making any further approximations.

### 2. Finite temperature $T < \infty$: exact wide-band evolution

The duality for the finite $T$ wide-band evolution (S-33) follows most clearly when working in an interaction-picture which is defined in the usual spirit by "extracting" the simple reference dynamics from the exact evolution:

$$\Pi(t) := \Pi_\infty(t)\, \Pi(t,0)_I. \tag{S-59}$$

Notably, in our case, this extracted evolution is already dissipative, as the reference solution (S-50) applies to the *coupled system in a certain physical limit*. In Eq. (S-59), the time-argument $(t,0)$ [instead of just $(t)$] serves to indicate that the interaction picture is defined with respect to the initial time 0 which will matter below. Since $\Pi_\infty(t)$ preserves self-adjointness for all $t$ [Eq. (S-51)], so does $\Pi(t,0)_I = \mathcal{K}\Pi(t,0)_I\mathcal{K}$. The derivation of the duality therefore reduces to first expressing $\Pi(t,0)_I^{\mathrm{T}} = [\mathcal{K}\Pi(t,0)_I\mathcal{K}]^\dagger$ in a dual evolution analogous to the $T = \infty$ case [Eq. (S-56)], and then transforming it back to the Schrödinger picture using

$$[\mathcal{K}\Pi(t)\mathcal{K}]^\dagger = [\mathcal{K}\Pi(t,0)_I\mathcal{K}]^\dagger \ [\mathcal{K}\Pi_\infty(t)\mathcal{K}]^\dagger = e^{-\Gamma t} [\mathcal{K}\Pi(t,0)_I\mathcal{K}]^\dagger \mathcal{P}\bar{\Pi}_\infty(t)\mathcal{P}. \tag{S-60}$$

Here, we used Eq. (S-48) and the $T = \infty$ duality (S-56).

In this interaction picture, the perturbation expansion for the wide-band evolution $\Pi(t,0)_I$ takes the familiar form:

$$\Pi(t,0)_I = (\mathbb{1}^{\mathrm{R}}|\mathcal{T}e^{-\int_0^t d\tau L_+^{\mathrm{T}}(\tau)}|\rho^{\mathrm{R}}) \tag{S-61a}$$

$$= \sum_{k=0}^\infty (-i)^k \int_{t \geq t_k \geq \ldots \geq t_1 \geq 0} dt_k \ldots dt_1 \, (\mathbb{1}^{\mathrm{R}}|L_+^{\mathrm{T}}(t_k) \ldots L_+^{\mathrm{T}}(t_1)|\rho^{\mathrm{R}}) \tag{S-61b}$$

$$= \sum_{\substack{k=0 \\ k \text{ even}}}^\infty (-i)^k \sum_{k'k\ldots 1'1} \int_{t \geq t_k \geq \ldots \geq t_1 \geq 0} dt_k \ldots dt_1 \, T_{k'k} \ldots T_{1'1} G_{k'}^+(t_k) \ldots G_{1'}^+(t_1)(\mathbb{1}^{\mathrm{R}}|J_k^-(t_k) \ldots J_1^-(t_1)|\rho^{\mathrm{R}}), \tag{S-61c}$$

where $\mathcal{T}$ denotes the time-ordering. In the last line, [b] Eq. (S-61c), the order $k$ is limited to even integers by the fermion-parity superselection [c]. In Eq. (S-61c), the Heisenberg field superoperators [d] with respect to $t = 0$ are defined in the standard way for the reservoirs, whereas for the subsystem, we use the exact $T = \infty$ *dissipative* dynamics (S-50):

$$J_l^q(t) := e^{iL^{\mathrm{R}}t} J_l^q \, e^{-iL^{\mathrm{R}}t}, \qquad G_{l'}^q(t) := \Pi_\infty(-t)\, G_{l'}^q \, \Pi_\infty(t). \tag{S-62}$$

The time-dependent perturbation expansion (S-61c) was established in Ref. [6], based on earlier Laplace-space formulations [5, 12]. This perturbation expansion is specific to the wide-band limit: it relies on the assumption that the couplings $T_{l'l}$ do not depend on the reservoir frequencies $\omega_l$ contained in the multi-index reservoir sums in Eq. (S-61c).

---

[a] The notation used in this supplement differs from that in Refs. [5, 6] where the overbar has a different meaning.

[b] To keep the notation simple, $k$ is an index on $T_{k'k}$ or $G_{k'}^+$ denotes the $k$-th multi-index, whereas $t_k$ is the $k$-th time argument. Similarly, the first sum is over all possible even values of $k$, while the second sum is a sum over multi-indices.

[c] It is not Wick's theorem which requires this: at this point we do not yet assume noninteracting electrodes.

[d] Here, the time argument suffices to indicate the interaction picture.

As discussed in Sec. II A 5, each such multi-index sum is a sum over all indices necessary to characterize the system. For the model considered in the main article, it explicitly reads

$$\sum_l = \sum_{\eta_l \sigma_l} \int_{-\infty}^{\infty} d\omega_l. \tag{S-63}$$

Since the fields of the local subsystem $G_{l'}^q(t)$ by definition [a] do not depend on reservoir frequencies $\omega_l$, *the reservoir frequency dependence in the wide-band limit is entirely contained in the reservoir fields.* [b] Exploiting this simple fact and furthermore using that *only one type of reservoir field ($J_l^-$) occurs in* (S-61c), the finite temperature duality in the dissipative interaction picture is found to take the elegant form

$$[\mathcal{K}\Pi(t,0)_I\mathcal{K}]^\dagger = \mathcal{P}\bar{\Pi}(0,-t)_I\mathcal{P} \quad , \quad \bar{\Pi}(0,-t)_I := \Pi(0,-t; \bar{L}, \bar{L}^{\mathrm{T}}, \bar{\mu})_I. \tag{S-64}$$

In addition to the dual quantities Eq. (S-58), this involves the dual chemical potentials:

$$\bar{\mu} := -\mu \tag{S-65}$$

We derive Eq. (S-64) from two basic observations: [c]

1. *The adjoint $\dagger$ combined with operator-adjoint $\mathcal{K} \bullet \mathcal{K}$ can be expressed entirely in terms of parity operations.* The established result Eq. (S-46) shows how this works for Schrödinger picture fields. Using the $T = \infty$ duality (S-56), we can extend this elegantly to interaction-picture fields (S-62):

$$[\mathcal{K}G_{1'}^q(t)\mathcal{K}]^\dagger = q\mathcal{P}(-\mathbb{1})^{L^N}\bar{G}_{1'}^q(-t)\mathcal{P}, \qquad \bar{G}_{l'}^q(t) := \bar{\Pi}_\infty(-t)\, G_{l'}^q\, \bar{\Pi}_\infty(t), \tag{S-66}$$

where on the right hand side, we now have $\bar{G}_{1'}^q$, the field $G_{1'}^q$ in the interaction picture with respect to the *dual* $T = \infty$ evolution [Eq. (S-56)] generated by $\bar{L}_\infty$ [Eq. (S-57)], as indicated by the overbar. Applying $[\mathcal{K} \bullet \mathcal{K}]^\dagger$ to a sequence of an even number $k$ of fields of the same type $q = +$, their order is reversed [Eq. (S-48)], their times are inverted, and a total sign $(-1)^{k/2}$ appears:

$$[\mathcal{K}G_{k'}^+(t_k)\dots G_{1'}^+(t_1)\mathcal{K}]^\dagger = (-\mathbb{1})^{L^N}\bar{G}_{1'}^+(-t_1)\dots(-\mathbb{1})^{L^N}\bar{G}_{k'}^+(-t_k) = (-1)^{k/2}\bar{G}_{1'}^+(-t_1)\dots\bar{G}_{k'}^+(-t_k). \tag{S-67}$$

The sign arises [d] by pairing up all parity superoperators $(-\mathbb{1})^{L^N}$ and using $[(-\mathbb{1})^{L^N}]^2 = +1$ [Eq. (S-46) ff.].

2. *Under the reservoir frequency integrals and reservoir average, we are free to simultaneously invert the ordering of the fields $J^-$, their time-arguments and all chemical potentials.* We can thus write

$$\sum_{k\dots1} F_{k\dots1}(\mathbb{1}^{\mathrm{R}}|J_k^-(t_k)\dots J_1^-(t_1)|\rho^{\mathrm{R}}) = \sum_{k\dots1} F_{k\dots1}\, (\mathbb{1}^{\mathrm{R}}|J_1^-(-t_1)\dots J_k^-(-t_k)|\rho^{\mathrm{R}})\big|_{\mu\to-\mu}, \tag{S-68}$$

where $F_{k\dots1}$ denotes any expression depending on all indices contained in the sequence of multi-indices $k \dots 1$ *except* for the frequencies $\omega_k, \dots \omega_1$. It requires two considerations to see this. First, inverting the ordering of the fields gives a sign, since fields of the same type $q = -$ anticommute [Eq. (S-39)], and the sign equals $(-1)^{k(k-1)/2} = (-1)^{k/2}$ for even $k$. Second, by inverting all times, we transform $J_l^q(t) = e^{-i\eta_l\omega_l t}J_l^q$ to $J_l^q(-t) = e^{-i\eta_l(-\omega_l)t}J_l^q$. The sign in the exponential function can be reverted by changing the sign of all frequency integration variables, which is possible since we integrate from $-\infty$ to $\infty$. However, the inverted frequency signs then also appear in the Wick expansion [e] of the correlator $(\mathbb{1}^{\mathrm{R}}|J_k^-\dots J_1^-|\rho^{\mathrm{R}})$, given by the sum of all permutations of pair contractions $\sum_{\mathrm{contr.}}(-1)^P \prod_{\langle n,l\rangle}(\mathbb{1}^{\mathrm{R}}|J_n^- J_l^-|\rho^{\mathrm{R}})$, since each Wick contraction depends on the frequency through [f]

$$(\mathbb{1}^{\mathrm{R}}|J_2^- J_1^-|\rho^{\mathrm{R}}) = \delta_{\bar{2}1}\tanh\left(\frac{\eta_2(\omega_2-\mu_2)}{2T_2}\right). \tag{S-69}$$

---

[a] Note that in perturbation theory renormalized by continuous RG transformations, the vertex operators $G$ do pick up a frequency dependence. For this reason, it is an interesting open question if our derivation can be modified to show that the RG preserves our duality or breaks it.

[b] Note that non-trivial dependence on all other reservoir quantum numbers is allowed for.

[c] Notably, the sign $(-1)^k$ appears *three times*, and thus survives. This sign leads to the imaginary factor in the duality $\bar{H}^{\mathrm{T}} = iH^{\mathrm{T}}$. The sign appears once in Eq. (S-67) and twice in the derivation of Eq. (S-68) where it cancels. Up to Eq. (S-67), all expressions are valid for arbitrary initial nonequilibrium reservoir state $\rho^{\mathrm{R}}$ (i.e. possibly interacting). It is only in the last step establishing Eq. (S-68) that we used that an equilibrium reservoir state $\rho^{\mathrm{R}}$ is determined by the noninteracting reservoir Hamiltonian $H^{\mathrm{R}}$. This crucial property of the noninteracting, equilibrium reservoirs ultimately allows the adjoint to be expressed in *forward* time-evolution of some dual model in the wide band limit.

[d] Schematically, $(-\mathbb{1})^{L^N}\bar{G}(-\mathbb{1})^{L^N}\bar{G} = (-\mathbb{1})^{2L^N}(-1)\bar{G}\bar{G} = -\bar{G}\bar{G}$.

[e] See Eq. (60) and Appendix A of Ref. [5].

[f] See Eq. (S-83) in App. II D 2 below and Ref. [6] [Eq. (101)].

Hence, to leave all pair contractions unaltered apart from an overall sign, the frequency inversion must be accompanied by the inversion of all chemical potentials $\mu$. Altogether, this yields

$$(\mathbb{1}^{\mathrm{R}}|J_2^-(t_2)J_1^-(t_1)|\rho^{\mathrm{R}}) = e^{-i(\eta_2\omega_2 t_2 + \eta_1\omega_1 t_1)}(\mathbb{1}^{\mathrm{R}}|J_2^- J_1^-|\rho^{\mathrm{R}}) = (-1)\,(\mathbb{1}^{\mathrm{R}}|J_2^-(-t_2)J_1^-(-t_1)|\rho^{\mathrm{R}})\big|_{\mu \to -\mu,\,\omega_l \to -\omega_l}. \tag{S-70}$$

For $k/2$ pair contractions (even $k$), one obtains a sign $(-1)^{k/2}$, which however cancels the sign due to the reordering of the reservoir fields in Eq. (S-68), and thereby verifies the latter equation.

The duality (S-64) follows immediately from the general structure of the wide-band limit perturbation expansion (S-61c) and the two observations pointed out above: taking the adjoint using Eq. (S-67) gives Eq. (S-71a) and corresponds – apart from a parity transformation $\mathcal{P}$ – to inverting all times as well as the order of the operators in which they appear. Using Eq. (S-68) to obtain Eq. (S-71b), we recover a *forward time-evolution*, starting at $-t$ and ending at time 0:

$$[\mathcal{K}\Pi(t,0)_I\mathcal{K}]^\dagger$$

$$= \sum_{\substack{k=0 \\ k \text{ even}}}^{\infty} (-i)^k \sum_{k'k\ldots1'1} \int_{t \geq t_k \geq \ldots \geq t_1 \geq 0} dt_k\ldots dt_1\, T_{k'k}\ldots T_{1'1}\mathcal{P}(-1)^{k/2}\bar{G}_{1'}^+(-t_1)\ldots\bar{G}_{k'}^+(-t_k)\mathcal{P}(\mathbb{1}^{\mathrm{R}}|J_k^-(t_k)\ldots J_1^-(t_1)|\rho^{\mathrm{R}}) \tag{S-71a}$$

$$= \mathcal{P}\sum_{\substack{k=0 \\ k \text{ even}}}^{\infty} (-i)^k \sum_{k'k\ldots1'1} \int_{0 \geq -t_1 \geq \ldots \geq -t_k \geq -t} dt_1\ldots dt_k\, T_{1'1}\ldots T_{k'k}i^k\bar{G}_{1'}^+(-t_1)\ldots\bar{G}_{k'}^+(-t_k)(\mathbb{1}^{\mathrm{R}}|J_1^-(-t_1)\ldots J_k^-(-t_k)|\bar{\rho}^{\mathrm{R}})\mathcal{P} \tag{S-71b}$$

$$= \mathcal{P}\sum_{k=0}^{\infty} (-i)^k \int_{0 \geq -t_1 \geq \ldots \geq -t_k \geq -t} dt_1\ldots dt_k\,(\mathbb{1}^{\mathrm{R}}|\bar{L}_+^{\mathrm{T}}(-t_1)\ldots\bar{L}_+^{\mathrm{T}}(-t_k)|\bar{\rho}^{\mathrm{R}})\mathcal{P} \tag{S-71c}$$

$$= \mathcal{P}(\mathbb{1}^{\mathrm{R}}|\mathcal{T}e^{-i\int_{-t}^0 d\tau \bar{L}_+^{\mathrm{T}}(-\tau)}|\bar{\rho}^{\mathrm{R}})\mathcal{P} = \mathcal{P}\bar{\Pi}(0,-t)_I\mathcal{P} \tag{S-71d}$$

Note that the dual time evolution $\bar{\Pi}$ is defined with respect to the same reservoir Liouvillian $L^{\mathrm{R}}$, but a dual reservoir *density operator* $\bar{\rho}^{\mathrm{R}} = \rho^{\mathrm{R}}|_{\mu \to -\mu}$ in Eq. (S-71b) with inverted chemical potentials.

Clearly, $\bar{\Pi}(0,-t)_I$ and $\bar{\Pi}(t,0)_I$ are related by a *time-translation (not: time-reversal!)*, since our total system is time-translation invariant. In the Schrödinger picture, the time evolution is equal in both cases, but in the interaction picture, a time-shift results in a shift of the reference time:

$$J_l^q(-t_l) = e^{iL^{\mathrm{R}}(-t_l)}J_l^q e^{-iL^{\mathrm{R}}(-t_l)} = e^{-iL^{\mathrm{R}}t_l}J_l^q(t - t_l)\,e^{iL^{\mathrm{R}}t_l}, \tag{S-72}$$

$$\bar{G}_{l'}^q(-t_l) = \bar{\Pi}_\infty(t_l)\,G_{l'}^q\,\bar{\Pi}_\infty(-t_l) = \bar{\Pi}_\infty(t)\,\bar{G}_{l'}^q(t-t_l)\,\bar{\Pi}_\infty(-t). \tag{S-73}$$

Since the transformation Eq. (S-72) drops out under the reservoir average in the step (S-71b), we obtain:

$$[\mathcal{K}\Pi(t,0)_I\mathcal{K}]^\dagger \overset{(\text{S-71d})}{=} \mathcal{P}\bar{\Pi}(0,-t)_I\mathcal{P} = \mathcal{P}\bar{\Pi}_\infty(t)\bar{\Pi}(t,0)_I\bar{\Pi}_\infty(-t)\mathcal{P}. \tag{S-74}$$

The Schrödinger picture dual evolution is defined, analogously to Eq. (S-59), by extracting the dual trivial $T = \infty$ evolution:

$$\bar{\Pi}(t) = \bar{\Pi}_\infty(t)\bar{\Pi}(t,0)_I. \tag{S-75}$$

Finally, inserting this and Eq. (S-74) into Eq. (S-60), and using preservation of hermiticity, $\Pi = \mathcal{K}\Pi\mathcal{K}$ [Eq. (S-47)], as well as $\mathcal{P}^2 = 1$, we arrive at

$$\Pi(t)^\dagger = [\mathcal{K}\Pi(t)\mathcal{K}]^\dagger = e^{-\Gamma t}\mathcal{P}\bar{\Pi}(t)\mathcal{P}. \tag{S-76}$$

This is the *general fermion-parity duality for the propagator in the time-representation*.

### 3. Kernel duality in frequency representation [Eq. (1)]

Equation (S-76) essentially completes the derivation. It remains to convert it [cf. Eq. (S-30)] into the equivalent kernel duality Eq. (1) reported in the main article, which is more convenient for applications. With our definition of

the Laplace transform, $f(\omega) = \int_0^\infty dt\, e^{i\omega t} f(t)$, applying $\int_0^\infty dt\, e^{i\omega^* t}$ to the duality (S-76) yields the propagator duality in the Laplace-frequency representation:

$$
[\Pi(\omega; L, L^{\mathrm{T}}, \mu)]^\dagger \overset{\text{Eq. (S-30)}}{=} \frac{-i}{\omega^* - \left[L + W(\omega; L, L^{\mathrm{T}}, \mu)\right]^\dagger}
$$

$$
\overset{\text{Eq. (S-76)}}{=} \frac{i}{\bar\omega - \mathcal{P}\left[\bar L + W(\bar\omega; \bar L, \bar L^{\mathrm{T}}, \bar\mu)\right]\mathcal{P}} \overset{\text{Eq. (S-30)}}{=} \mathcal{P}\,\Pi(\bar\omega; \bar L, \bar L^{\mathrm{T}}, \bar\mu)\,\mathcal{P}. \tag{S-77}
$$

The second line is the transform of $\mathcal{P}\bar\Pi(t)\mathcal{P}$ at shifted complex frequency $\bar\omega = -\omega^* + i\Gamma$. Using Eq. (S-30) and $\bar L = -L$, we can thus directly read off the duality between the kernels, the central result Eq. (1) of the main article:

$$
\left[W(\omega; L, L^{\mathrm{T}}, \mu)\right]^\dagger = -\Gamma + \mathcal{P}W(\bar\omega; \bar L, \bar L^{\mathrm{T}}, \bar\mu)\mathcal{P} \tag{S-78}
$$

In the main article, the functional dependence on the Hamiltonian was written. Here, this dependence enters through $\bar L = [\bar H, \bullet]$, and $\bar L^{\mathrm{T}} = [\bar H^{\mathrm{T}}, \bullet]$.

### C. Duality as a consistency check on calculations [Eq. (1)]

As mentioned in the discussion of the duality (1) in the main article, the right hand side of Eq. (S-78) *can always be calculated* if the kernel $W$ has been obtained, *either exactly or approximately*, as a function of the parameters specifying the model, i.e., the parameters appearing in $H$, $H^{\mathrm{T}}$, and the $\mu$'s in $\rho^{\mathrm{R}}$. One simply substitutes these parameters into $W(\omega; H, H^{\mathrm{T}}, \mu)$ such that $H \to -H$ and $H^{\mathrm{T}} \to iH^{\mathrm{T}}$, respectively, and inverts all $\mu$'s in this *known* function to obtain $W(\bar\omega; \bar H, \bar H^{\mathrm{T}}, \bar\mu)$. For example, for the Anderson model in the wide-band limit with spin-dependent tunneling, $\Gamma = \sum_\sigma \Gamma_\sigma$, the kernel duality reads: [a]

$$
\left[W(\omega; \epsilon, U, \Gamma_\sigma, \mu)\right]^\dagger = -\Gamma + \mathcal{P}W(i\Gamma - \omega^*; -\epsilon, -U, -\Gamma_\sigma, -\mu)\mathcal{P}. \tag{S-79}
$$

Since we have proven the duality in a general context, it can thus be *exploited* for any model that satisfies our assumptions [see Sec. II A 5] when performing some perturbative or nonperturbative calculation of the kernel.

Conversely, it is easy to check explicitly if a calculation actually preserves the exact relation Eq. (1). For example, we have verified that in the limit of *vanishing interactions* but for *arbitrary coupling* $\Gamma$ relative to $T$, the duality is satisfied by the exact, wide-band limit solution for the nonequilibrium, non-Markovian dynamics of an arbitrary multilevel Anderson model. The exact reduced density operator for this problem, i.e., including all multiparticle correlations on the subsystem (which factorize in this limit), was computed in Ref. [6]. The result for the exact propagator $\Pi(t)$, written in the time representation, was written in a compact form that most explicitly satisfies the duality Eq. (S-76) when the above mentioned substitutions are performed.

### D. Collection of some relevant derivations

To keep this supplement self-contained, we finally write out and comment on some of the algebra leading to the key formula (S-46) for converting adjoints into parity operations, and to the reservoir correlator (S-69). We emphasize that these results were already established in Ref. [5] and are rederived here for the reader's convenience. However, since they provide the crucial starting point for the novel derivation in Sec. II B of Eq. (1) of the main article, we here calculate them in a form that relies less on the context of Ref. [5].

#### 1. Superoperator adjoint [Eq. (S-40)], operator-adjoint [Eq. (S-42)], and parity [Eq. (S-45)]

Relation (S-46) was obtained by simply composing (S-40),(S-42) and (S-45). The latter three relations all follow directly from the definition of the field superoperators (S-37).

---

[a] Note that the couplings also feature in the frequency argument through their lumped sum $\Gamma = \sum_\sigma \Gamma_\sigma$: here, the $\Gamma_\sigma$'s do not get inverted (they are not model parameters here, but part of the imaginary frequency shift).

*a. Proof of Eq. (S-40) (adjoint):* For arbitrary operators $A$ and $B$, the adjoint $(G_2^q)^\dagger$ can be written as

$$(A|(G_2^q)^\dagger B) \overset{(1)}{=} (G_2^q A|B) \overset{(2)}{=} \tfrac{1}{\sqrt{2}} \operatorname{Tr}\left[\{d_2 A + q(-\mathbb{1})^N A(-\mathbb{1})^N d_2\}^\dagger B\right] \tag{S-80}$$

$$= \tfrac{1}{\sqrt{2}} \operatorname{Tr}\left[\{A^\dagger d_2^\dagger + q d_2^\dagger (-\mathbb{1})^N A^\dagger (-\mathbb{1})^N\} B\right] \overset{(3,4)}{=} \tfrac{1}{\sqrt{2}} \operatorname{Tr}\left[A^\dagger \{d_2 B - q(-\mathbb{1})^N B(-\mathbb{1})^N d_2\}\right] = (A|G_2^{\bar q} B).$$

We used (1) the definition of the adjoint of superoperator $S$, $(A|S^\dagger B) := (SA|B)$; (2) the definition of the field superoperator, Eq. (S-37); (3) the definition of the overbarred multi-index $\bar 2$, $d_{\bar 2}^\dagger = d_2$ [Eq. (S-41) ff.]; (4) the cyclic trace property and, in the second term, the property $d_2(-\mathbb{1})^N = -(-\mathbb{1})^N d_2$ of the fermion-parity operator. The proof for $(J_1^q)^\dagger$ is equivalent.

*b. Proof of Eq. (S-42) (operator-adjoint):* Consider the action on an arbitary operator $|B) = B$:

$$\mathcal{K}\left(G_2^q(\mathcal{K}B)\right) \overset{(1)}{=} \mathcal{K}(G_2^q B^\dagger) \overset{(2)}{=} \tfrac{1}{\sqrt{2}}\mathcal{K}\left[d_2 B^\dagger + q(-\mathbb{1})^N B^\dagger(-\mathbb{1})^N d_2\right] \overset{(3)}{=} \tfrac{1}{\sqrt{2}}\left[B d_2 + q d_2(-\mathbb{1})^N B(-\mathbb{1})^N\right]$$

$$\overset{(4)}{=} \tfrac{1}{\sqrt{2}}\left[(-\mathbb{1})^N B'(-\mathbb{1})^N d_2 + q d_2 B'\right] \overset{(5)}{=} q G_2^q B' \overset{(6)}{=} q G_2^q (-\mathbb{1})^{L^N} B. \tag{S-81}$$

Here, we employed: (1) the definition of $\mathcal{K}$, $\mathcal{K}B = B^\dagger$; (2) the definition of the field superoperator Eq. (S-37); (3) the definition of $\mathcal{K}$ and $\mathcal{K}(AB) = B^\dagger A^\dagger = (\mathcal{K}B)(\mathcal{K}A)$. (4) We temporarily set $B' = (-\mathbb{1})^N B(-\mathbb{1})^N = (-\mathbb{1})^{L^N} B$ and thus $B = (-\mathbb{1})^N B'(-\mathbb{1})^N$, (5) extract the sign $q$ using $q = \pm 1$ to recover the definition Eq. (S-37), (6) and finally restore $B$ again.

*c. Proof of Eq. (S-45) (parity transform):* Consider the action on $|B) = B$, where $B$ is an arbitrary operator:

$$\mathcal{P}(G_2^q)\mathcal{P}B \overset{(1)}{=} (-\mathbb{1})^N\left\{G_2^q\left\{(-\mathbb{1})^N B\right\}\right\} \overset{(2)}{=} \tfrac{1}{\sqrt{2}}(-\mathbb{1})^N\left\{d_2(-\mathbb{1})^N B + qB(-\mathbb{1})^N d_2\right\}$$

$$\overset{(3)}{=} \tfrac{1}{\sqrt{2}}\left[-d_2 B + q(-\mathbb{1})^N B(-\mathbb{1})^N d_2\right] \overset{4}{=} -G_2^{\bar q} \text{ with } \bar q := -q. \tag{S-82}$$

We applied (1) the definition of $\mathcal{P}$ as left-multiplication by the parity operator, $\mathcal{P}\bullet = (-\mathbb{1})^N\bullet$, (2) the property $[(-\mathbb{1})^N]^2 = \mathbb{1}$, (3) and the property $d_2(-\mathbb{1})^N = -(-\mathbb{1})^N d_2$ of the fermion-parity.

### 2. Reservoir correlator (S-69)

Finally, we write out the calculation of the single (!) reservoir correlation function (S-69) that appears in expansion (S-61c) of the wide-band limit propagator (S-33). Although its expression is well-known [a], we write out the calculation to show how it arises from the definition of the fields (S-37). Inserting this definition, one calculates Eq. (S-69) as follows:

$$\operatorname{Tr}_{\mathrm{R}} J_2^- J_1^- \rho^{\mathrm{R}} \overset{(1)}{=} \operatorname{Tr}_{\mathrm{R}} \tfrac{1}{\sqrt{2}} J_2^- \left\{[c_1, \rho^{\mathrm{R}}]_+\right\} \overset{(2)}{=} \tfrac{1}{2}\operatorname{Tr}_{\mathrm{R}}[c_2, [c_1, \rho^{\mathrm{R}}]_+]_- \overset{(3)}{=} \operatorname{Tr}_{\mathrm{R}}[c_2, c_1]_- \rho^{\mathrm{R}}$$

$$\overset{(4)}{=} \delta_{\bar 2 1} \operatorname{Tr}_{\mathrm{R}}\left[\delta_{\eta_2, +}[c_{\sigma_2 \omega_2}^\dagger, c_{\sigma_2 \omega_2}]_- + \delta_{\eta_2, -}[c_{\sigma_2 \omega_2}, c_{\sigma_2 \omega_2}^\dagger]_-\right] \rho^{\mathrm{R}} \overset{(5)}{=} \delta_{\bar 2 1} \eta_2 \operatorname{Tr}_{\mathrm{R}}\left(2c_{\sigma_2 \omega_2}^\dagger c_{\sigma_2 \omega_2} - 1\right)\rho^{\mathrm{R}}$$

$$\overset{(6)}{=} \delta_{\bar 2 1} \eta_2 \left(2f((\omega_2 - \mu)/T) - 1\right) \overset{(7)}{=} \delta_{\bar 2 1} \tanh\left(\frac{\eta_2(\omega_2 - \mu_2)}{2T}\right). \tag{S-83}$$

Here, we used the following simple facts: (1) the definition of fields (S-37) and that the reservoir state – by the superselection postulate – must have even electron parity, $(-\mathbb{1})^{N^{\mathrm{R}}} \rho^{\mathrm{R}}(-\mathbb{1})^{N^{\mathrm{R}}} = \rho^{\mathrm{R}}$, i.e., it has no unpaired electron operators; (2) one "excitation" creates odd-parity operators, $(-\mathbb{1})^{N^{\mathrm{R}}}[c_1, \rho^{\mathrm{R}}]_+(-\mathbb{1})^{N^{\mathrm{R}}} = -[c_1, \rho^{\mathrm{R}}]_+$ using $(-\mathbb{1})^{N^{\mathrm{R}}} c_1 = -c_1(-\mathbb{1})^{N^{\mathrm{R}}}$; (3) the trace is cyclic; (4) the expectation is nonzero only if $1 = \bar 2 = (-\eta_2)\sigma_2 \omega_2$ and $2 = \eta_2 \sigma_2 \omega_2$ with $\eta_2 = \pm$; (5) commutators are antisymmetric; (6) $\operatorname{Tr}_{\mathrm{R}} c_{\sigma_2 \omega_2}^\dagger c_{\sigma_2 \omega_2} \rho^{\mathrm{R}} = f(x) = (e^x - 1)^{-1}$ with $x = (\omega_2 - \mu_2)/T$; (7) $2f(x) - 1 = \tanh(x/2)$ and $\eta_2 \tanh(x/2) = \tanh(\eta_2 x/2)$ for $\eta_2 = \pm$.

[1] M. Leijnse, M. R. Wegewijs, and M. H. Hettler, Phys. Rev. Lett. **103**, 156803 (2009).

---

[a] After steps (3) and (4) in Eq. (S-83), one recognizes that Eq. (S-69) is just the Keldysh Green function of the noninteracting reservoir. See Ref. [5], p. 5-7, for an explanation how the field superoperators (S-37) exploit the advantages of the Keldysh formalism in the density-operator framework.

[2] T. Fujisawa, D. G. Austing, Y. Tokura, Y. Hirayama, and S. Tarucha, J. Phys. Cond. Mat. **15**, R1395 (2003).

[3] C. Rössler, T. Krähenmann, S. Baer, T. Ihn, K. Ensslin, C. Reichl, and W. Wegscheider, New J. Phys. **15**, 033011 (2013).

[4] A. Beckel, A. Kurzmann, M. Geller, A. Ludwig, A. D. Wieck, J. König, and A. Lorke, EPL (Europhysics Letters) **106**, 47002 (2014).

[5] R. B. Saptsov and M. R. Wegewijs, Phys. Rev. B **86**, 235432 (2012).

[6] R. B. Saptsov and M. R. Wegewijs, Phys. Rev. B **90**, 045407 (2014).

[7] H.-P. Breuer and F. Petruccione, *The Theory of Open Quantum Systems* (Oxford University Press, 2002).

[8] S. Mukamel, *Principles of nonlinear optical spectroscopy* (Oxford University Press, Oxford, 1995).

[9] C. Timm, Phys. Rev. B **77**, 195416 (2008).

[10] S. Koller, M. Grifoni, M. Leijnse, and M. R. Wegewijs, Phys. Rev. B **82**, 235307 (2010).

[11] J. König, J. Schmid, H. Schoeller, and G. Schön, Phys. Rev. B **54**, 16820 (1996).

[12] H. Schoeller, Eur. Phys. Journ. Special Topics **168**, 179 (2009).

[13] H. Schoeller and F. Reininghaus, Phys. Rev. B **80**, 045117 (2009).

[14] G. C. Wick, A. S. Wightman, and E. P. Wigner, Phys. Rev. **88**, 101 (1952).

[15] Y. Aharonov and L. Susskind, Phys. Rev. **155**, 1428 (1967).

[16] R. Streater and A. Wightman, *PCT, spin and statistics, and all that*, Landmarks in Physics (Princeton University Press, 2000).

[17] N. N. Bogolubov, A. A. Logunov, A. I. Oksak, and I. T. Todorov, *General principles of quantum field theory* (Nauka, Moscow, 1987) in Russian. English translation: Kluwer Academic Publishers, Dordrecht, 1989.


\end{document}